\documentclass[10pt, a4paper]{article}
\pdfoutput=1
\usepackage{jheppub}
\usepackage[utf8]{inputenc}
\usepackage{amsmath}
\usepackage{amsfonts}
\usepackage{amssymb}
\usepackage{physics}
\usepackage{float}
\usepackage{caption}
\usepackage{subcaption}
\usepackage{ulem}
\newcommand{\OO}{\mathcal{O}}
\newcommand{\tfdl}{\langle \text{TFD}|}
\newcommand{\tfdr}{|\text{TFD}\rangle}

\newmuskip\pFqmuskip
\newcommand*\pFq[6][8]{%
  \begingroup % only local assignments
  \pFqmuskip=#1mu\relax
  \mathcode`\,=\string"8000
  \begingroup\lccode`\~=`\,
  \lowercase{\endgroup\let~}\pFqcomma
  {}_{#2}F_{#3}{\left[\genfrac..{0pt}{}{#4}{#5};#6\right]}%
  \endgroup
}

\newcommand{\pFqcomma}{\mskip\pFqmuskip}

\begin{document}
\title{A Cardy formula for off-diagonal three-point coefficients; or, how the geometry behind the horizon gets disentangled}
\author{Aurelio Romero-Berm\'udez,}
\author{Philippe Sabella-Garnier,}
\author{Koenraad Schalm}
\affiliation{Instituut-Lorentz for Theoretical Physics, $\Delta ITP$, Leiden University, Niels Bohrweg 2, Leiden 2333CA, The Netherlands}
\emailAdd{romero@lorentz.leidenuniv.nl}
\emailAdd{garnier@lorentz.leidenuniv.nl}
\emailAdd{kschalm@lorentz.leidenuniv.nl}

\abstract{In the AdS/CFT correspondence eternal black holes can be
  viewed as a specific entanglement between two copies of the CFT: 
the thermofield double. The statistical CFT
  Wightman function can be computed from a geodesic between the two
  boundaries of the Kruskal extended black hole and therefore probes
  the geometry behind the horizon. We construct a kernel
  for the AdS$_3$/CFT$_2$ Wightman function that is independent of the
  entanglement. This kernel equals the average off-diagonal matrix element
  squared of a primary operator. This allows us to compute the
    Wightman function for an arbitrary entanglement between the double
    copies and probe the emergent geometry
  between a left- and right-CFT that are not thermally entangled.}

\maketitle

\section{Introduction}
The holographic mapping between the two-sided black hole in Anti de Sitter space (AdS) and the thermofield double (TFD) state of a strongly-coupled conformal field theory \cite{Maldacena:2001kr} is one of the most useful settings for the understanding of quantum gravity. It allows for the reformulation of questions concerning the black hole interior in terms of unitary quantum mechanics involving both copies of the boundary field theory. In this context, the black hole entropy is accounted for by the entropy associated to the entanglement between the two boundaries. This can be understood in the bulk as a consequence of the Ryu-Takayanagi formula \cite{Ryu:2006bv}: the minimal surface homologous to one of the boundaries is located at the horizon.

In \cite{VanRaamsdonk:2010pw}, the idea of using the Ryu-Takayanagi formula to understand how entanglement ``stitches up'' the bulk geometry was introduced. More precisely, a thought experiment was proposed: suppose the two boundaries are progressively disentangled. What would happen to the bulk geometry? The Ryu-Takayanagi formula tells us that the area of the minimal surface separating the two boundaries (i.e. the black hole horizon) would decrease. The geodesic approximation, 
\begin{equation}
e^{-\Delta \mathcal{L}(x_1,x_2)} \sim \langle \mathcal{O}(x_1)
\mathcal{O}(x_2)\rangle
\end{equation}
and the relationship between mutual information and correlation functions tells us that the {proper} distance between the two boundaries would increase. These two observations lead to the conclusion that the bulk would pinch off. This general picture can be checked by examining the geometry of black holes of different masses, since decreasing the mass leads to a decrease of entanglement. A bulk calculation was also presented in \cite{Czech:2012be}. The idea of probing the geometry behind the horizon using two-point functions was also extensively studied in \cite{Balasubramanian:1999zv,Fidkowski:2003nf,Festuccia:2005pi}.

In this paper, we make the answer to the question of what happens to
the geometry when entanglement is decreased more precise. It builds on
the observation that the two-sided black hole is more appropriately
viewed as dual to thermal Schwinger-Keldysh path-integral
\cite{Herzog:2002pc} and the correlation function between operators
inserted on opposite boundaries dual to geodesics for large conformal
dimension computes the Wightman function
\cite{Festuccia:2005pi}. The Wightman function vanishes for $T=0$,
where there is no entanglement in the TFD description, and one
directly infers that at $T=0$ the two Schwinger-Keldysh copies of the
CFT geometrically disconnect. Clearly Feynman Green's
  functions dual to operators inserted on one side remain
  well-defined, showing that it is only the geometry inside the
  horizon that changes drastically while the geometry outside the horizons stays
  smooth as one changes $T$.  Computing the Wightman function in the
TFD for an arbitrary entanglement between the two copies that interpolates between
no entanglement and thermal thus allows one to probe in some detail the emergence of
the geometry behind the horizon. This is the aim of this paper.

 Focusing on
AdS$_3$/CFT$_2$,  we do so and explicitly calculate boundary two-point functions
between scalar primary operators inserted on the two boundaries for
states which have less entanglement than the TFD. A crucial step
in this computation is our construction of % we
% \new{first} find
an expression for the average  matrix element squared of a scalar primary operator $\mathcal{O}$ of dimension $\Delta$ at given energies:
\begin{equation}
\mathcal{F}(E_a,E_b,\Delta)\equiv\overline{|\langle a| \mathcal{O} | b\rangle|^2} =e^{-S(E_a)} e^{-S(E_b)} \sum_{\substack{a|E(a)=E_a \\ b|E(b)=E_b}} |\langle a |\mathcal{O}|b\rangle|^2~, \label{defcurlyF}
\end{equation}
where $e^{S(E)}$ is the number of states at energy $E$. \footnote{Note that the states being averaged over are \emph{all} the energy eigenstates at a given energy, not just primary states. Using radial quantization, this expression is clearly related, through torus one-point conformal blocks, to the average of the square of OPE coefficients of primary operators.} This can be seen as a generalization of the expression found for
average diagonal elements found in \cite{Kraus2017a}. We obtain this
formula by inverting the relationship between ${\cal
    F}(E_a,E_b,\Delta)$ and the two-point function of primary
  operators inserted on both boundaries in the thermofield double
\begin{equation}
\tfdl \OO(t_1) \otimes \OO(t_2) \tfdr = \frac{1}{Z(\beta)} \sum_{E_a, E_b} e^{-\beta {E_a+E_b \over 2}} e^{-i (t_2-t_1)(E_a-E_b)} \mathcal{F}(E_a,E_b,\Delta)~,
\end{equation}
using the fact that the 
holographic expression for the left-hand side is
  known. As we were completing this article, we were alerted that
  Brehm, Das, Datta also computed  $\mathcal{F}(E_a,E_b,\Delta)$ by
  different means \cite{Brehmetal}. Similar work was also subsequently done by Hikida, Kusuki, Takayanagi \cite{Hikida:2018khg}. Moreover, an average over heavy primaries, based on  large-$c$ conformal blocks, has been found in  \cite{Cardy2017}. 
The resulting expression for ${\cal
    F}(E_a,E_b,\Delta)$ reflects properties of eigenstate
thermalization.
For high enough $\Delta$ in non-integrable CFTs we expect the eigenstate thermalization hypothesis (ETH) to hold \cite{ETH1,ETH2,ETH3,ETHreview}:
\begin{equation}
\langle a |O |b \rangle= O(E_a) \delta_{a b} + e^{-\frac{1}{2}S\left(\frac{E_a+E_b}{2}\right)}f(E_a,E_b)R_{ab}~, \label{eq:eth}
\end{equation}
where $O(E)$ is the average of the diagonal matrix elements at energy
$E$, $f(E_a,E_b)$ is a smooth function and $R_{ab}$ is a random number
with vanishing average and mean-squared equal to unity.
In particular ${\cal
    F}(E_a,E_b,\Delta)$ is closely related to the variance function
  $f(E_a,E_b)$, as we shall show.

With our construction of an expression for ${\cal
    F}(E_a,E_b,\Delta)$ and the density of states, we
can compute Wightman functions for states with entanglement patterns
different than that of the thermofield double. The simplest such
pattern is one which brutally interpolates between effective thermal
entanglement below an arbitrary hard cut-off $\Lambda$ and {\em zero}
entanglement above this cut-off. 
\begin{equation}
|\Lambda\rangle \equiv \frac{1}{\sqrt{Z_\Lambda}} \sum_{a : E_a \leq \Lambda} e^{-\beta E_a/2} |a\rangle |a\rangle~,
\qquad Z_\Lambda \equiv  \sum_{a : E_a \leq \Lambda} e^{-\beta E_a/2}
\end{equation}
For this set of non-thermally, lesser entangled ``doubled'' states
  we examine the dependence of the Wightman function on time
separation and cutoff. They clearly
  illustrate that erasing entanglement destroys a conventional notion
  of a smooth connected geometric spacetime behind the horizon.

\section{Constructing ${\cal
    F}(E_a,E_b,\Delta)$: the average of the square of matrix elements}\label{sec:Setup}
\subsection{Setup}
Our starting point is the global AdS$_3$ black hole. This is
dual to a two-dimensional CFT defined on a spatial circle of size $L$
at finite inverse temperature $\beta$. Thermal correlation functions in
  this CFT can be computed from the thermofield double state, where the latter
  equals by definition
\begin{equation}
\tfdr=\frac{1}{\sqrt{Z(\beta)}} \sum_a e^{-\beta E_a/2} |a\rangle |a\rangle~.
\end{equation}
Here the sum is to be taken over all energy eigenstates $|a\rangle$ of a single copy of the CFT. The one-sided Hamiltonian $H$ is such that $H|a\rangle=E_a|a\rangle$. $Z(\beta)$ is the usual thermal partition function,
\begin{equation}
Z(\beta)=\sum_a e^{-\beta E_a}~,
\end{equation}
which ensures that the state is properly normalized. 
We consider time-evolution in the thermofield double with the
  Hamiltonian
\begin{equation}
H_\text{tot}=1\otimes H - H \otimes 1~, 
\end{equation}
so that the whole state is stationary. Therefore, for an operator inserted on the left boundary we have
\begin{align}
\OO(t,x)\otimes 1 &= e^{iH_\text{tot} t} (\OO(x) \otimes 1 )e^{-iH_\text{tot} t}  \nonumber \\
&= (e^{-i H t} \OO(x) e^{i H t})\otimes 1~,
\end{align}
matching the notion that ``time runs backwards'' on one of the
boundaries.\footnote{Note that we will stick with this convention for
  the direction of time. Operators on the left boundary should always
  be thought of as being time-evolved with the two-sided Hamiltonian,
  so that if they are at a positive value of $t$ they are inserted in
  the bottom half of the Penrose diagram. This is because, with our
  choice of Hamiltonian, real time on either boundary matches the real
  part of the Schwarzschild time coordinate in the bulk in the
  appropriate patch. In other words, the symmetry associated with the
  bulk Killing vector consists in shifting the time coordinate on both
  boundaries by the same constant.} 

The expectation value of the TFD
two-point function --- dual to operators inserted on opposite
boundaries --- is
related to the thermal Wightman function by analytic
continuation. Assuming that $\OO$ is Hermitian
\begin{align}\label{eq:TFD_correlator}
\tfdl \OO(t_1,\phi_1) \otimes \OO(t_2,\phi_2) \tfdr &= \frac{1}{Z(\beta)} \sum_{a,b} e^{-\beta \frac{E_a+E_b}{2}} \langle a |e^{-i H t_1} \OO(\phi_1) e^{i H t_1} |b \rangle \langle a| e^{i H t_2} \OO(\phi_2) e^{-i H t_2} |b\rangle \nonumber \\
&=\frac{1}{Z(\beta)} \text{Tr}(\rho_\beta \OO(t_1-i\beta/2,\phi_1) \OO(t_2,\phi_2))~,
\end{align}
where $\rho_\beta$ is the usual thermal density matrix. In what
follows, we will simplify the calculation by taking both operators to
be inserted at the same coordinate $\phi$.\footnote{In Appendix \ref{app1}, we generalise the results of this section to the case of operators inserted at different spatial coordinates $\phi_1 \neq \phi_2$. } Since they are on opposite
boundaries (or, if analytically continued to a one-sided calculation,
separated in time by at least $i\beta/2$), we do not have to worry
about issues of ordering or contact divergences.

Substituting the definition of the TFD one has
\begin{align}\label{eq:TFD_correlator_2}
 \!\!\!G^{\beta}(t_2-t_1)\equiv \tfdl \OO(t_1) \otimes \OO(t_2) \tfdr &= \frac{1}{Z(\beta)} \sum_{a,b} e^{-\beta (E_a+E_b)/2} e^{-i (E_a-E_b)(t_2-t_1)} |\langle a |\OO  |b \rangle|^2~% \\
% &~. \nonumber
\end{align}
We can split the sum over states into a sum over energies and a sum of states at that energy:
\begin{align}
G^{\beta}(t_2-t_1) &= \frac{1}{Z(\beta)} \sum_{E_a, E_b} e^{-\beta(E_a+E_b)/2-i(E_a-E_b)(t_2-t_1)} \sum_{\substack{a|E(a)=E_a \\ b|E(b)=E_b}} |\langle a|\OO|b\rangle|^2 \nonumber\\
&=\frac{1}{Z(\beta)} \sum_{E_a, E_b} e^{-\beta(E_a+E_b)/2-i(E_a-E_b)(t_2-t_1)} e^{S(E_a)} e^{S(E_b)} \mathcal{F}(E_a,E_b,\Delta)~.
\end{align}
Replacing the weighted sum over energy levels with an integral
  over the density of states $\rho(E_{a,b})$, \footnote{
In the large $c$ limit, or in a holographic CFT, the spectrum of a 2D
CFT becomes continuous, making this approximation exact.}
\begin{equation}
\sum_{E_{a},E_{b}} e^{S(E_a)} e^{S(E_b)} \rightarrow \int dE_a dE_b \rho(E_a) \rho(E_b)~, 
\end{equation}
and making
the following change of variables:
\begin{equation}
E=\frac{E_a+E_b}{2}-E_L ~~,~~ \chi=E_a-E_b ~~,~~ t=t_2-t_1~
\end{equation}
with $E_L=-\frac{\pi c}{6L}$ being the Casimir energy,
we obtain an expression for the TFD two-point function in terms of the
average of the square of matrix elements ${\cal
    F}(E_a,E_b,\Delta)$, as defined in \ref{defcurlyF}.
\begin{align}
G^{\beta}(t) &= \frac{1}{Z(\beta)} \int\limits_{E_L}^\infty \dd E_a\  \rho(E_a) \int\limits_{E_L}^\infty \dd E_b\  
\rho(E_b)\   e^{-\beta{E_a+E_b\over 2}}\ e^{-i(E_a-E_b)t} \mathcal{F}(E_a,E_b,\Delta) \nonumber \\
&=\frac{e^{-\beta E_L}}{Z(\beta)}
\int\limits_{-\infty}^{\infty} \dd \chi e^{-i\chi t} 
\int\limits_{0}^{\infty} \dd Ee^{-\beta E}
\rho\left(E+{\chi\over 2}\right)\rho\left(E-{\chi\over 2}\right) 
 \Theta\left(E{-}{|\chi|\over 2}\right)\mathcal{F}(E,\chi,\Delta)~, \label{defG}
\end{align}
where $\Theta$ is a step function enforcing the obvious bound $\chi \leq 2E$.

We now use the fact that the LHS is known from holography in the large
$c$ limit \cite{Maldacena:2001kr}.
The thermal Wightman function in the deconfined state dual
  to a black hole equals\footnote{At finite $c$ and on a finite
    interval $L$ there is a unique Wightman function, as there are no
    phase transitions in 1+1 dimensions at finite volume. Eq.\eqref{confG} is a
    truncation of this exact result in the limit of large $c$. We
    shall also mostly consider the infinite volume limit $L\rightarrow\infty$.}
\begin{equation}
G^{\beta}(t_1-t_2)=\sum_{n\in \mathbb{Z}} C_\OO \frac{(2\pi/\beta)^{2\Delta}}{\left[\cosh \left(2\pi n L/\beta\right)+\cosh\left(2\pi (t_1-t_2)/\beta\right)\right]^\Delta}~, \label{confG}
\end{equation}
where ${C}_\OO$ is a normalization constant and the sum over
images ensures spatial periodicity $\phi \simeq \phi + 2\pi n$
-- recall that both operators are inserted at the same spatial
coordinate $\phi =0$. 
Note that as $L \rightarrow \infty$, only the $n=0$ term
survives. This is also the case if we take $\Delta \rightarrow
\infty$. In holography, the $n\neq0$ images correspond to winding
geodesics, which should be suppressed.

Matching Equations \eqref{defG} and \eqref{confG} gives us:
\begin{align}
&Z(\beta)e^{\beta E_L}\sum_{n\in \mathbb{Z}} C_\OO
  \frac{(2\pi/\beta)^{2\Delta}}{\left[\cosh \left(2\pi n
  L/\beta\right)+\cosh\left(2\pi t/\beta\right)\right]^\Delta}~
= \nonumber\\
&~~~~
\int\limits_{-\infty}^{\infty} \dd \chi e^{-i\chi t} 
\int\limits_{0}^{\infty} \dd Ee^{-\beta E}
\rho\left(E+{\chi\over 2}\right)\rho\left(E-{\chi\over 2}\right) 
 \Theta\left(E{-}{|\chi|\over 2}\right)\mathcal{F}(E,\chi,\Delta)~. 
\label{eq:bigF_formal}
\end{align}
Notice that the left-hand side is nothing more than a Fourier
transform in $\chi$ followed by a Laplace transform in $E$. Inverting these two transforms will give us an expression for $\mathcal{F}(E,\chi,\Delta)$. We now proceed to do so. 
\subsection{Inverting the Fourier transform}\label{sec:FT}
Inverting the Fourier transform on the RHS of
Eq.\eqref{eq:bigF_formal} is straightforward. For the LHS we
use the following integral identity
\begin{align}
\int\limits_{-\infty}^\infty \dd x{e^{i k x}\over [b+\cosh(x)]^\Delta}&={2^{\Delta-1}\over \Gamma(\Delta)}
\left[
\left|\Gamma\left({\Delta-ik \over2}\right)\right|^2
{}_2F_1\left({\Delta+ik \over2},{\Delta-ik \over2},\frac{1}{2},b^2\right)\right.\notag\\
&\left.-2b\left|\Gamma\left({\Delta+1-ik \over2}\right)\right|^2
{}_2F_1\left({\Delta+1+ik \over2},{\Delta+1-ik \over2},\frac{3}{2},b^2\right)\right]\,.
\end{align}
This gives us
\begin{align}
  \label{eq:formal}
  \int_{-\infty}^{\infty} \dd t e^{i\chi t} 
%Z(\beta)e^{\beta E_L} C_\OO 
\sum_{n\in \mathbb{Z}}
  \frac{(2\pi/\beta)^{2\Delta}}{\left[\cosh \left(2\pi n
  L/\beta\right)+\cosh\left(2\pi t/\beta\right)\right]^\Delta}~ =
%Z(\beta) e^{\beta E_L} C_{\OO} 
\frac{1}{{2\pi}} \left(\frac{2\pi}{\beta}\right)^{2\Delta} \sum_{n \in \mathbb{Z}} F_n(\beta,\chi,\Delta)\,,
\end{align}
where
\begin{align}
b_n=&\cosh(2\pi n L/\beta) \\
F_n(\beta,\chi,\Delta) 
=&{\beta\over 2\pi}{2^{\Delta-1}\over \Gamma(\Delta)} 
\left[
\left|\Gamma\left({\Delta-i {\beta\chi\over 2\pi}\over2}\right)\right|^2
{}_2F_1\left(
{\Delta+i {\beta\chi\over 2\pi}\over2},
{\Delta-i {\beta\chi\over 2\pi}\over2},
\frac{1}{2},b_n^2\right)\right.\notag\\
&\qquad 
\left.-2b_n
\left|\Gamma\left({\Delta+1-i {\beta\chi\over 2\pi}\over2}\right)\right|^2
{}_2F_1\left(
{\Delta+1+i {\beta\chi\over 2\pi}\over2},
{\Delta+1-i {\beta\chi\over 2\pi}\over2},{3\over 2},
b_n^2\right)\right]~.\label{eq:inverseFT}
\end{align}
Note that $F_n$ is a smooth function of $\beta$ and $\chi$ and is even in $\chi$. Furthermore, $F_n(\beta=0,\chi,\Delta)=0$ and $F_n(\beta,\chi=0,\Delta)\propto \beta$. For the special case of $n=0$, we have
\begin{align} \label{eq:F_line}
F_{n=0}(\beta,\chi,\Delta) = &
{\beta\over 2\pi} \int\limits_{-\infty}^{\infty} \dd t {e^{-i t {\chi \beta\over 2\pi}}\over [ 1+\cosh(t)]^\Delta} = 
{\beta\over 2\pi} {2^{\Delta}\over \Gamma(2\Delta)} 
\Gamma\left(\Delta-i{\chi\beta\over 2\pi}\right)
\Gamma\left(\Delta+i{\chi\beta\over 2\pi}\right)\,.
\end{align}
The identity Eq.\eqref{eq:bigF_formal} after inverting the Fourier
transform becomes:
\begin{equation}\label{eq:inverseLT}
\int\limits_{0}^{\infty} \dd E e^{-\beta E}
\rho\left(E+{\chi\over 2}\right)\rho\left(E-{\chi\over 2}\right) 
 \Theta\left(E{-}{|\chi|\over 2}\right)\mathcal{F}(E,\chi,\Delta)=
Z(\beta) e^{\beta E_L} \frac{{C}_O}{{2\pi}} \left(\frac{2\pi}{\beta}\right)^{2\Delta} \sum_{n \in \mathbb{Z}} F_n(\beta,\chi,\Delta)\,.
\end{equation}

\subsection{Inverting the Laplace transform}\label{sec:LT}
We now use the Mellin inversion theorem to invert the Laplace transform. This inversion theorem states that, given a function $H(z)$ analytic in $\mathbb{C}\setminus \{z_1,\dots,z_n\}$ (the $z_i$ being isolated singularities), its inverse Laplace transform is
\begin{equation}
h(w) = \mathcal{L}^{-1}\left[H(z);w\right] = {1\over 2\pi i }\int\limits_{\gamma_0+i\mathbb{R}} \dd z\  e^{z w} H(z) = \sum_{j=1}^n \text{Res}(e^{z w}H(z);z_j)\,, \ \ \gamma_0 > \Re(z_i)\,.
\end{equation}
The integral is taken over a vertical line in the complex
$z$ plane which is set to the right of all the singularities of
$H(z)$. We then use the residue theorem by closing the contour to the
left (the so-called `` Bromwich contour''). It turns out to be simplest
to split the inversion by finding the inverse Laplace transform of
$e^{\beta E_L} Z(\beta)$ and of
$\left(\frac{2\pi}{\beta}\right)^{2\Delta} F_n(\beta,\chi,\Delta)$
separately and convolving them. Going from the $\beta$ domain to the $E$ domain, we have:
\begin{align}
\label{eq:doe}
\rho(E+E_L)\equiv& \mathcal{L}^{-1} \left[e^{\beta E_L}Z(\beta);E\right] ~, \\
\hat{F}_n(E,\chi,\Delta)\equiv& \mathcal{L}^{-1} \left[\left(\frac{2\pi}{\beta}\right)^{2\Delta} F_n(\beta,\chi,\Delta);E\right]~.\label{eq:Fhat} 
\end{align}
The identity determining ${\cal F}(E,\chi,\Delta)$ in terms of these functions is
\begin{equation}
\rho\left(E+{\chi\over 2}\right)\rho\left(E-{\chi\over 2}\right) 
 \Theta\left(E{-}{|\chi|\over 2}\right)\mathcal{F}(E,\chi,\Delta) = 
 \frac{C_\OO}{{2\pi}}\int\limits_{0}^{E} \dd E' \rho(E'{+}E_L) \sum_{n \in \mathbb{Z}}\hat{F}_n(E-E',\chi,\Delta)~.\label{eq:curlyF}
\end{equation}

Restricting ourselves to integer $\Delta>1$, we compute $\hat{F}_n(E,\chi,\Delta)$ as defined in Eq.\eqref{eq:Fhat} with the inverse Mellin transform. We observe that the combination
$\left(\frac{2\pi}{\beta}\right)^{2\Delta}\!\! F_n(\beta,\chi,\Delta)$ has
simple poles at:
\begin{align}
\beta=\pm \beta_\ell^{(1)}\equiv& \pm 2\pi i \frac{2\ell+\Delta}{\chi}~,\\
\beta=\pm \beta_\ell^{(2)}\equiv& \pm 2\pi i \frac{2\ell+\Delta+1}{\chi}~,\quad \ell \in \mathbb{N}
\end{align}
 and a pole of order $2\Delta-1$ at $\beta=0$. Summing over the
   residues at these poles, we find that
\begin{equation}
\begin{split}
\hat{F}_n(E,\chi,\Delta)&=
2\pi(-1)^\Delta{ 2^{\Delta+1}\over \Gamma(\Delta)}
 \chi^{2\Delta-2} \times\\
&\sum_{\ell=0}^\infty
{(-1)^{\ell}\over \ell!}
\left[
{\cos(2\pi(2\ell+\Delta){E-E_L\over\chi})\over (2\ell+\Delta)^{2\Delta-1}}
{\Gamma(\Delta+\ell) }\ 
{}_2F_1\left(-\ell,\ell+\Delta,{1\over2},b_n^2\right)\right.\\
&-2b_n\left. 
{\cos(2\pi(2\ell+\Delta+1){E-E_L\over\chi})\over (2\ell+\Delta+1)^{2\Delta-1}}
{\Gamma(\Delta+\ell+1)}\ 
{}_2F_1\left(-\ell,\ell+\Delta+1,{3\over2},b_n^2\right)
\right]\\
&+ \text{Res}\left[e^{\beta E}\left({2\pi\over \beta}\right)^{2\Delta} {F}_n(\beta,\chi,\Delta);\beta = 0\right]~.\label{eq:Fnhat}
\end{split}
\end{equation}

Next we compute $\rho(E+E_L)$ defined in Eq.\eqref{eq:doe}.
This function is the density of states. Recall that $E_L=-\frac{\pi
  c}{6L}$ is the Casimir energy.
By definition,
\begin{align}
Z(\beta)&= \int_{E_L}^\infty d\tilde{E} e^{-\beta \tilde{E}} \rho(\tilde{E})
\end{align}

Shifting the range of integration one has
\begin{align}
Z(\beta) = e^{-\beta E_L} \int_0^\infty dE e^{-\beta E} \rho(E+E_L)~.
\end{align}
Thus $\rho(E+E_L)$ indeed equals
  the density of states.
To compute it, we need to inverse Laplace
transform the partition function. In \cite{Hartman:2014oaa}, it was shown that, to leading order in $c$, the partition function in the deconfined phase is given by
\begin{equation}
Z(\beta)=e^{\frac{\pi c}{6} \frac{L}{\beta}}~.
\end{equation}
Thus, using the Mellin inversion formula (for $E+E_L>0$), we have
\begin{equation}
\rho(E+E_L)=\text{Res}\Big[e^{\beta E} e^{\beta E_L +\frac{\pi c}{6} \frac{L}{\beta}};\beta=0\Big]~,
\end{equation}
since the only singularity is the essential singularity at $\beta=0$. We can write the Laurent series to evaluate the residue:
\begin{equation}
e^{\beta(E+E_L)}e^{\frac{\pi c L}{6\beta}}=\sum_{k=0}^{\infty} \sum_{m=0}^\infty \frac{(E+E_L)^k}{k!} \frac{1}{m!}\left(\frac{6\beta}{\pi c L}\right)^{-m}~.
\end{equation}
The residue at $\beta=0$ corresponds to the $k=m-1$ term, hence
\begin{align}
\rho(E+E_L)&=\sum_{m=0}^\infty \frac{(E+E_L)^{m-1}}{m! (m-1)!} \left(-L^2 E_L\right)^m \nonumber\\
&=L \frac{\sqrt{-E_L}}{\sqrt{E+E_L}} I_1\left(2L\sqrt{-E_L}\sqrt{E+E_L}\right)~, \label{doseq}
\end{align}
where $I_1$ is a modified Bessel function. In terms of the usual
energy coordinate $\tilde{E}=E+E_L \in \left[-\frac{\pi c}{6L},\infty
\right]$ and substituting the explicit expression for the Casimir
energy $E_L<0$, this equals
\begin{equation}
\label{eq:DoS}
\rho(\tilde{E})=\sqrt{\frac{\pi c}{6}} \sqrt{\frac{L}{\tilde{E}}} I_1\left(2 \sqrt{\frac{\pi c}{6}} \sqrt{L \tilde{E}}\right)
\end{equation}
This inherently $c\gg 1$
expression is also only valid for $\tilde{E}>0$,
i.e. $E>-E_L=|E_L|=\frac{\pi c}{6L}$, since
we can only close the Bromwich contour to the left when $e^{\beta
  (E+E_L)}$ vanishes for $\text{Re}(\beta)=-\infty$. For values
  $-E_L>E>E_L$ the Mellin transform as it stands appears ill-defined, but
  this must be cured by all lower-order in $c$ corrections to the
partition function, which are not universal.\footnote{See more comments on these corrections in App. \ref{app:DoS}.} We shall only consider
$E\gg |E_L|$. Then, keeping $c$ fixed and taking $\tilde{E} L \rightarrow \infty$, we recover the Cardy formula:
\begin{equation}
\rho(\tilde{E}) = \frac{1}{2} \left(\frac{cL}{6 \pi
    \tilde{E}^3}\right)^{1/4} e^{2\sqrt{\frac{\pi c}{6} \tilde{E} L}}
+ \ldots~.
\label{cardy}
\end{equation}

\subsection{The limit $L\rightarrow \infty$: decompactifying the CFT
  from the circle to the line.}
Equations \eqref{eq:Fnhat} and \eqref{doseq} give a formal solution to
$\mathcal{F}(E,\chi,\Delta)$. However, the fact that this solution is
expressed in terms of two infinite sums --- one over images from the
  periodic boundary condition on the interval $L$, see
  Eq. \eqref{eq:curlyF}, and one over infinite poles,
  Eq. \eqref{eq:Fnhat} --- is an obstacle to explicitly evaluating
  correlation functions in our modified state. 
We will therefore study the system with the size $L$ of the
  system to be large. This way we discard one of the sums and restrict
  ourselves to only the $n=0$ image in Eq. \eqref{eq:curlyF}.
In that regime $E_L \rightarrow 0$, so we are justified in ignoring 
the non-universal part of the spectrum mentioned above.

Restricting to only the $n=0$ term, it turns out the remaining
  sum over poles can be formally done. One can show that the identity for $\mathcal{F}(E,\chi,\Delta)$ reduces to
\begin{equation}\label{eq:F_definition}
\rho\left(E+{\chi\over 2}\right)\rho\left(E-{\chi\over 2}\right) \mathcal{F}(E,\chi,\Delta)=\frac{{C}_\OO}{{2\pi}} \int_0^E \rho(E') \hat{F}_0(E'-E,\chi,\Delta) dE'
\end{equation}
with 
\begin{equation}
\begin{split}
\hat{F}_0(E,\chi,\Delta)&=2\pi{(-1)^\Delta 2^{\Delta}\over \Delta^{2\Delta-1}}
\chi^{2\Delta-2} 
  \times\\
&\left\{
e^{ 2\pi i \Delta E/\chi}\ 
\pFq[1]{2\Delta}{2\Delta-1}{a_1,\dots,a_{2\Delta}}{b_1,\dots,b_{2\Delta-1}}{-e^{2\pi i E/\chi}} +
e^{- 2 \pi i \Delta E/\chi}\
\pFq[1]{2\Delta}{2\Delta-1}{a_1,\dots,a_{2\Delta}}{b_1,\dots,b_{2\Delta-1}}{-e^{-2\pi i E/\chi}}\right\} \\
&+{2^{\Delta} (2\pi)^{2\Delta-1} \over \Gamma(2\Delta)(2\Delta-2)!}{\dd^{2\Delta-2} \over\dd \beta^{2\Delta-2}} 
 \left.e^{\beta E}\Gamma\left({\Delta+i {\beta\chi\over 2\pi}}\right)\Gamma\left({\Delta-i {\beta\chi\over 2\pi} }\right)\right|_{\beta = 0} \\
&a_i={\Delta}~~ \text{for}~~ i\leq 2\Delta-1~~,~~ a_{2\Delta}=2\Delta~~ \text{and} ~~ b_i=1+{\Delta} ~~ \text{for}~~ i\leq 2\Delta-1~, \label{eq:Fhat_line} 
\end{split}
\end{equation}
and $\rho(E)$ as in Eq.\eqref{eq:DoS}.
Here, $\textstyle\pFq{2\Delta}{2\Delta-1}{a_1,\dots,a_{2\Delta}}{b_1,\dots,d_{2\Delta-1}}{y}$ is the generalised hypergeometric function with the series representation
\begin{equation}
\pFq{p}{q}{a_1,\dots,a_{p}}{b_1,\dots,d_{q}}{y} = 
\sum_{n=0}^\infty{(a_1)_n\dots(a_p)_n \over(b_1)_n\dots(b_q)_n}{y^n\over n!}~,
\end{equation}
and $(a)_n \equiv a(a+1)\dots (a+n-1) = {\Gamma(a+n)\over \Gamma(a)}$ being the Pochhammer symbol. 

We should comment that strictly speaking neither side has a well-defined
  decompactification limit $L\rightarrow \infty$, but diverges.
  However, it
 is easy to see that under a scale transformation $L\rightarrow
   \lambda L$ and $E\rightarrow E/\lambda,~\chi \rightarrow
   \chi/\lambda$ the function $\hat{F}_0(E,\chi,\Delta) \rightarrow
   \lambda^{2\Delta-2} \hat{F}_0(E,\chi,\Delta)$, and therefore
\begin{equation}
\mathcal{F}(E, \chi, \Delta) \xrightarrow[{L\rightarrow \lambda L}]{} \lambda^{2\Delta} \mathcal{F}(\lambda^{-1} E, \lambda^{-1} \chi,\Delta)~.
\end{equation}
Thus, as long as we keep the dimensionless variables $EL$ and $\chi L$ fixed and divide by the appropriate power of $L$, we can think of our expressions as being valid in a decompactifying limit $L\rightarrow \infty$. In that limit, the $n\neq 0$ winding terms in the Wightman function are suppressed. Since the transforms we are doing are linear, we can neglect the images and focus on the $n=0$ term as $L\rightarrow \infty$.\footnote{Additionally, if we take $\Delta$ to be large then the $n=0$ term dominates even at finite L.}

\section{Analysis of matrix elements and implications for ETH}
\subsection{Asymptotic formulae}

In the previous section, we have given an exact analytical
  expression for $\mathcal{F}(E,\chi,\Delta)$ for any integer
  $\Delta$, an arbitrary value of $\chi L$ at large $EL$ and 
  large $c$. 
However, there are a few approximations we can do to get asymptotic expressions that are more transparent.
To do so it is convenient to take one step back before we perform the
Mellin transform. Formally inverting Eq.~\eqref{eq:inverseLT} in the
decompactification limit where we only keep the term $n=0$, we have
\begin{align}
\mathcal{F}(E,\chi,\Delta)&= \frac{{C}_\OO}{2\pi}\frac{1}{\rho\left(E+{\chi\over 2}\right)\rho\left(E-{\chi\over 2}\right)} \frac{1}{2\pi i} \oint d\beta ~ e^{\beta E} Z(\beta) \left(\frac{2\pi}{\beta}\right)^{2\Delta}  F_0(\beta,\chi,\Delta)~, \notag \\
F_0(\beta,\chi,\Delta) &= {\beta\over 2\pi} {2^{\Delta}\over \Gamma(2\Delta)} 
\Gamma\left(\Delta-i{\chi\beta\over 2\pi}\right)
\Gamma\left(\Delta+i{\chi\beta\over 2\pi}\right)\,,
\label{eq:2}
\end{align}
where the integral is taken over the Bromwich contour. 

The most obvious thing to do is a saddle-point approximation of the integral, in the regime where $ELc\gg1$. In that regime, $e^{\beta E} Z(\beta)$ is strongly peaked around $\beta_0=\sqrt{\frac{\pi c L}{6E}}$. Including the Gaussian corrections, this leads to
\begin{equation}
\mathcal{F}(E,\chi,\Delta) \simeq \frac{2^{3\Delta-2}}{\pi \Gamma(2\Delta)}\frac{\rho(E)}{\rho(E+\chi/2)\rho(E-\chi/2)} \left(\frac{6E\pi}{cL}\right)^{\Delta-1/2} \Gamma\left(\Delta-\frac{i\chi}{2\pi}\sqrt{\frac{\pi c L}{6E}}\right) \Gamma\left(\Delta+\frac{i\chi}{2\pi}\sqrt{\frac{\pi c L}{6E}}\right)~.
\end{equation}
This approximation converges to the correct value in the large $E$ limit for any $\chi/E$, but does so at different rates since the width of peak in the integrand depends on $\chi$. It can therefore be used to get a good approximation of $\mathcal{F}(E,\chi=a E,\Delta)$ as $E\rightarrow \infty$ for fixed $\chi/E=a$, but should not be used to examine the $\chi/E=a$ dependence of that expression.
Note that doing this same saddle-point approximation for $\oint d\beta e^{\beta E} Z(\beta)$ recovers the Cardy formula with the correct prefactor, given in equation \eqref{cardy}, as seen in \cite{Carlip:2000nv}. This approach is similar to the one used in \cite{Kraus2017a} to obtain the average of the diagonal matrix elements of $\OO$. 

Alternatively, we can expand $F_0(\beta,\chi,\Delta)$ in
$\chi$ first, i.e. $\beta\chi, \chi L, \frac{\chi}{E} \ll 1$ and then
do the inverse Laplace transform exactly, term by term. For $ \chi\ll \{\beta^{-1},L^{-1},E\}$, 
\begin{align}
\left(\frac{2\pi}{\beta}\right)^{2\Delta} F_0(\beta,\chi,\Delta) = \left({2\pi\over \beta}\right)^{2\Delta-1}  {2^{\Delta}\Gamma\left(\Delta\right)^2 \over \Gamma(2\Delta)}
\left[1-{\psi^{(1)}(\Delta)\over 4\pi^2} (\beta\chi)^2+O\left((\beta\chi)^4\right)\right]   \, ,
\end{align}
where $\psi^{(1)}(\Delta)$ is the Polygamma function of order
one. Performing then the contour integral using residue calculus
  in the identity
Eq.\eqref{eq:2} gives:
\begin{align}
&\mathcal{F}(E,\chi,\Delta)=
\frac{{C}_\OO(2\pi)^{2\Delta-2}}{\rho\left(E+{\chi\over 2}\right)\rho\left(E-{\chi\over 2}\right)}{2^{\Delta}\Gamma\left(\Delta\right)^2 \over \Gamma(2\Delta)} 
{\left(6E\over \pi cL\right)}^{\Delta-1}
 \left[
 I_{2\Delta-2}\left(2\sqrt{E L \pi c\over 6}\right)\right.\notag\\
&\hspace{5cm}
\left. -\chi^2{\left(6E\over \pi cL\right)}^{-1} {\psi^{(1)}(\Delta)\over 4\pi^2} I_{2\Delta-4}\left(2\sqrt{E L \pi c\over 6}\right)+O(\chi^4 L^2/E^2)
 \right]\label{eq:Fchi}~,
\end{align}
where $I_{n}$ is the modified Bessel function of order $n$.
Note that all the subleading terms in this expansion are all of the
form: $a_N(\Delta)\chi^{2N}{\left(6E\over \pi cL\right)}^{-N}
{\psi^{(1)}(\Delta)\over 4\pi^2} I_{2\Delta-2N}\left(2\sqrt{E L \pi
    c\over 6}\right)$ for $N=2,3,4,\dots$, and the constants
$a_N(\Delta)$ involve Polygamma functions of higher order  depending
only on $\Delta$, e.g., the term shown in the above equation has
$a_1(\Delta)={\psi^{(1)}(\Delta)\over 4\pi^2}$. This is a good approximation to examine the $\chi$ dependence for a fixed value of $\chi/E\ll1$. Furthermore, as $EL\rightarrow
  \infty$, this approximation overlaps with the saddlepoint
  approximation, provided $\chi$ is held fixed and does not scale with
$E$. 
Indeed, when $\chi=0=E_a-E_b$, i.e. for the average of the squared matrix elements of the same energy, this approximation becomes exact, and we have
\begin{equation}
\mathcal{F}(E,\chi=0,\Delta)=\frac{{C}_\OO(2\pi)^{2\Delta-2}}{\rho(E)^2}{2^{\Delta}\Gamma\left(\Delta\right)^2 \over \Gamma(2\Delta)} 
{\left(6E\over \pi cL\right)}^{\Delta-1} I_{2\Delta-2}\left(2\sqrt{E L \pi c\over 6}\right)~. \label{eq:Fchi0}
\end{equation}

\subsection{Does $\mathcal{F}(E,\chi,\Delta)$ reflect eigenstate thermalization?}

\begin{figure}[H]
  \begin{subfigure}{0.5\textwidth}
    \centering{\includegraphics[scale=0.8]{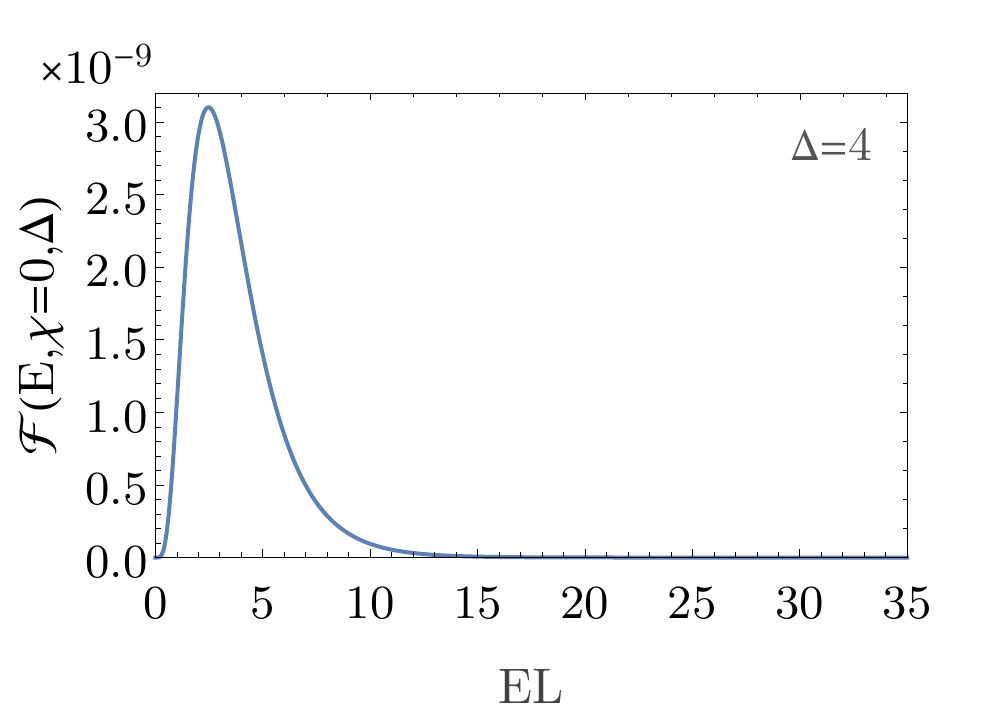}}
    %\caption{Caption text 1}
  \end{subfigure}
  \begin{subfigure}{0.5\textwidth}
    \centering{\includegraphics[scale=0.8]{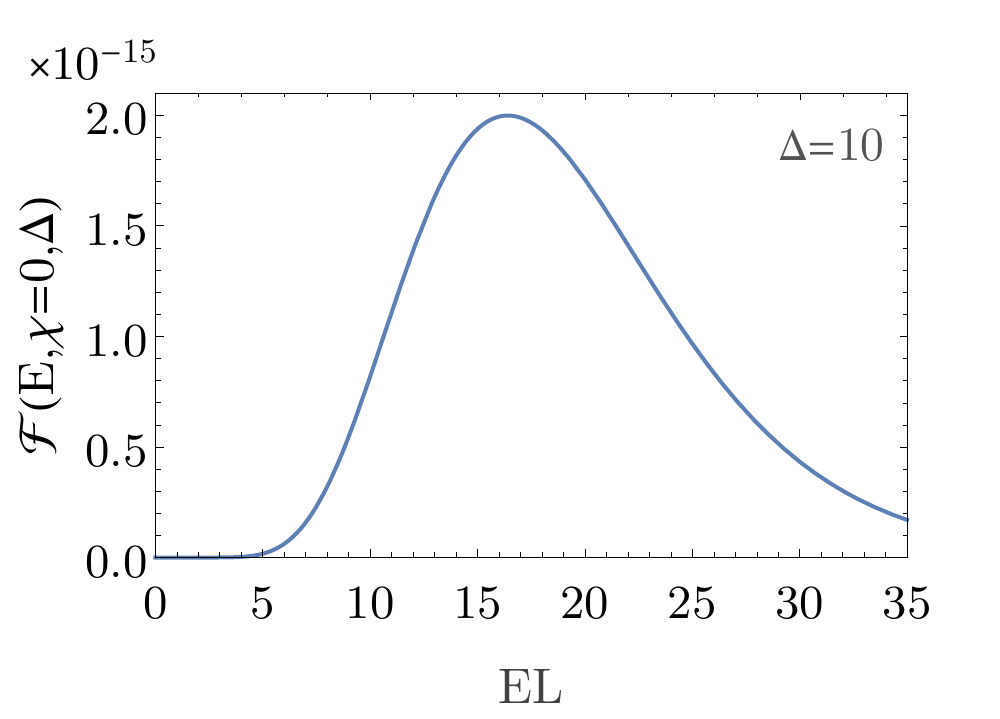}}
    %%\caption{Caption text 2}
  \end{subfigure}\\[-4ex]

  \begin{subfigure}{0.5\textwidth}
    \centering{\includegraphics[scale=0.8]{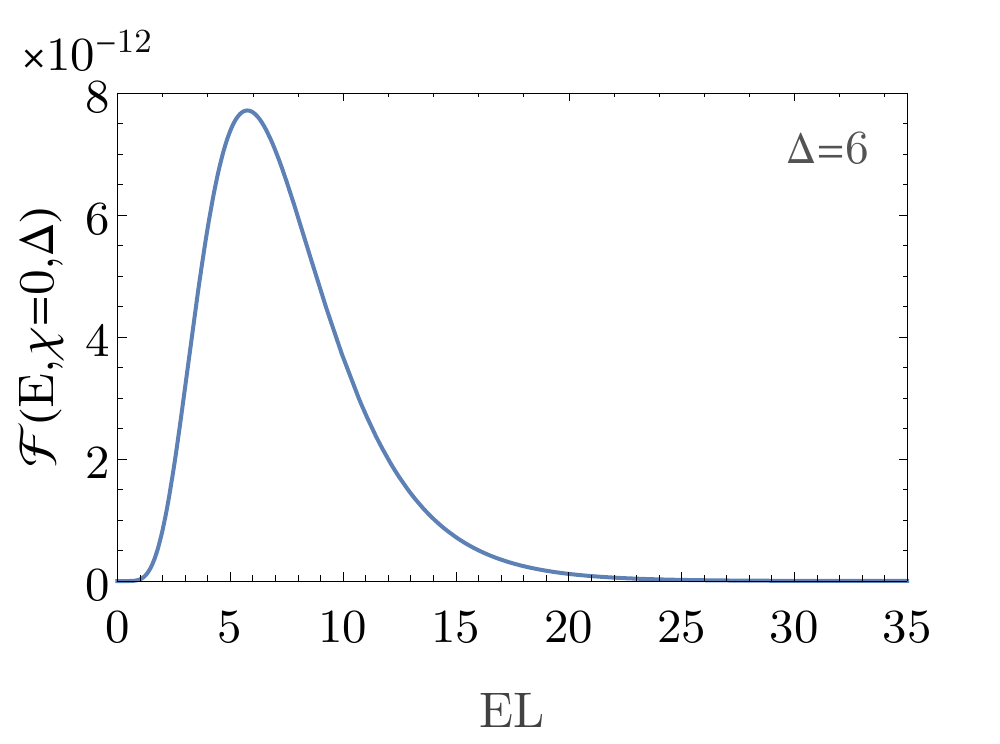}}
    %\caption{Caption text 1}
  \end{subfigure}
  \begin{subfigure}{0.5\textwidth}
    \centering{\includegraphics[scale=0.8]{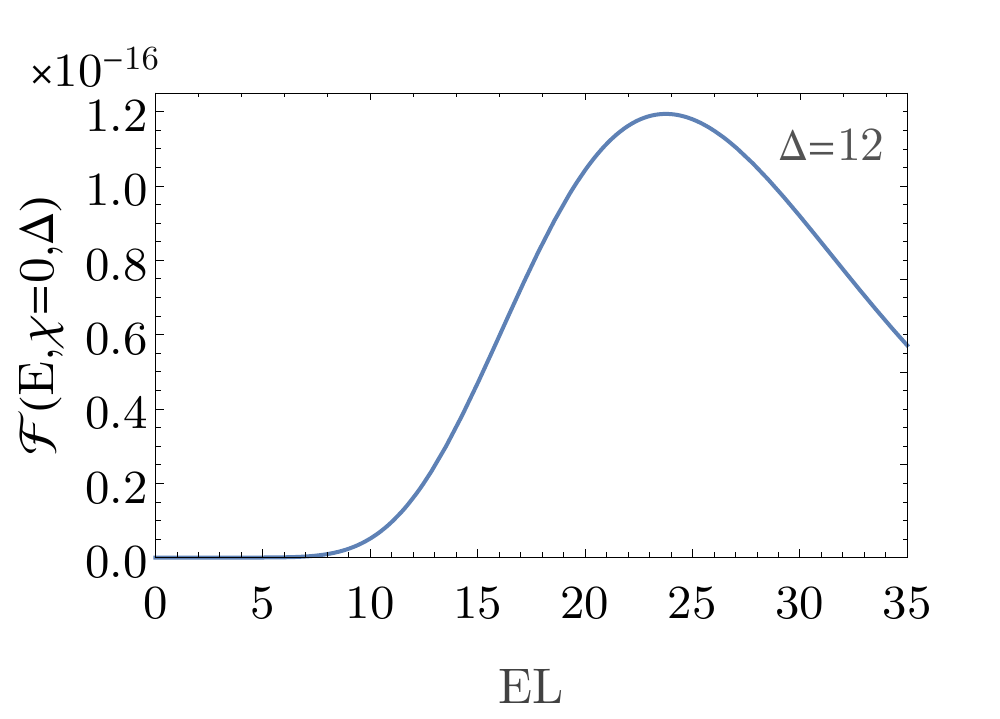}}
    %\caption{Caption text 2}
  \end{subfigure}
  \caption{Average of the square of matrix elements $\mathcal{F}(E,\chi=0,\Delta)$ of a scalar primary of dimension $\Delta$ as a function of the energy $E$ of the states (for states with the same energy), obtained from Eq. \eqref{eq:Fchi0} with $\chi=0$ and  $c=20$.}
	\label{fig:diag}
\end{figure}

It is instructive to compare the asymptotic results for
  $\mathcal{F}(E,\chi,\Delta)$ obtained in the previous subsection
  to the results predicted by eigenstate thermalization: the near thermal form
  of probe correlation functions in a closed many-body quantum system. 
 There are more and more
  hints that the actual physics underlying black hole thermodynamics
  is this eigenstate thermalization, rather than conventional
  dynamical classical thermalization (see e.g. \cite{Cotler:2016fpe,Lashkari2016}). 
Suppose that ETH is indeed at work. Then the mean-square of an
operator expectation value is well approximated by
\begin{align}
|\langle a |O| b \rangle|^2&= |O(E_a)|^2 \delta_{ab} +2e^{-S(E_a)/2} f(E_a,E_a)R_{ab} \delta_{ab}  + e^{-S(\frac{E_a+E_b}{2})} |f(E_a,E_b)|^2 |R_{ab}|^2~, \label{eq:avsq}
\end{align}

which is obtained by simply squaring (\ref{eq:eth}).\footnote{We focus here on spinless primary states for simplicity. More generally, we would expect states to organize in a generalized Gibbs ensemble. However, in the large $c$ limit, one expects to recover the usual Gibbs ensemble \cite{deBoer:2016bov,Basu:2017kzo,Lashkari:2017hwq}.}
Note that $a,b$ are not (yet) summed over. Substituting this in the
definition of $\mathcal{F}(E_a,E_b,\Delta)$, one sees that it is now
completely determined by the density of states $\rho(E)=e^{-S(E)}$,
the microcanonical value of the operator $|O(E_a)|$ and the universal variance
function $f(E_a,E_b)$:

\begin{align}
\mathcal{F}(E_a,E_b,\Delta)&=e^{-S(E_a)} e^{-S(E_b)} \left(|O(E_a)|^2 \delta_{E_a,E_b} \sum_{ab} \delta_{a,b} +  e^{-S(\frac{E_a+E_b}{2})} |f(E_a,E_b)|^2 \sum_{a,b} |R_{ab}|^2 \right) \nonumber \\
&= e^{-S(E_a)} e^{-S(E_b)} \left(  |O(E_a)|^2  e^{S(E_a)} \delta_{E_a,E_b} +  e^{-S(\frac{E_a+E_b}{2})} |f(E_a,E_b)|^2 e^{S(E_a)} e^{S(E_b)} \right) %\nonumber 
%&\leq|O(E_a)|^2 e^{-S(E_a)} \delta_{E_a,E_b} + e^{-S(\frac{E_a+E_b}{2})} |f(E_a,E_b)|^2~, 
\end{align}
where the sums over $a$ and $b$ are restricted to states with energy
$E_a$ and $E_b$ respectively. The term linear in $R_{ab}$ does not
appear since it averages to zero. 
At $\chi=0$, ETH predicts that
\begin{equation}
\mathcal{F}(E,\chi=0,\Delta) = e^{-S(E)} \left(|O(E)|^2 + |f(E,\chi=0)|^2\right)~.
\end{equation}
A priori, this includes equal contributions from diagonal matrix elements (the $O(E)$ term) and off-diagonal matrix elements of the same energy (the $f(E,\chi=0)$ term). In particular, note that while the off-diagonal terms were suppressed by a factor of the density of states in Eq. (\ref{eq:avsq}), they overcame that suppression in the average by their sheer number. Furthermore, we expect that $O(E)$ is in fact small compared to $f(E,\chi=0)$ since it is dual to the thermal one-point function of a free field (which vanishes at tree level). In \cite{Kraus2017b}, the one-loop correction was obtained. Using that result and  the exact expression for $\mathcal{F}(E,\chi=0,\Delta)$, Eq. \eqref{eq:Fchi0}, it is easy to see that, so long as $E/c$ is finite as $c\rightarrow \infty$,
\begin{equation}
\frac{|O(E)|}{|f(E,\chi=0)|} \leq A |\langle \mu |\OO |\mu \rangle|~,
\end{equation}
where $|\mu\rangle$ is the lightest state making the right-hand side non-zero and $A$ depends on $c$ only through $E/c$. $\langle \mu |\OO |\mu \rangle$ decays as $c^{-1}$ or faster, as we would expect when comparing a one-loop effect to a tree-level result \cite{Kraus2017b}.
The suppression of any discontinuities at $\chi=0$ (the result at $\chi\neq0$ does not include a contribution from $O(E)$) implies that the natural way to view $\mathcal{F}(E,\chi=0,\Delta)$ in holographic theories is as the variance of $O(E)$.
Using equation \eqref{eq:Fchi0} it is easy to see that $\mathcal{F}(E,\chi=0,\Delta)$ is peaked around $Ec\sim\Delta^2$. This is shown in figure \ref{fig:diag}.

\subsubsection{Matrix elements for different energies}
\begin{figure}[H]
    \centering{\includegraphics[scale=0.5]{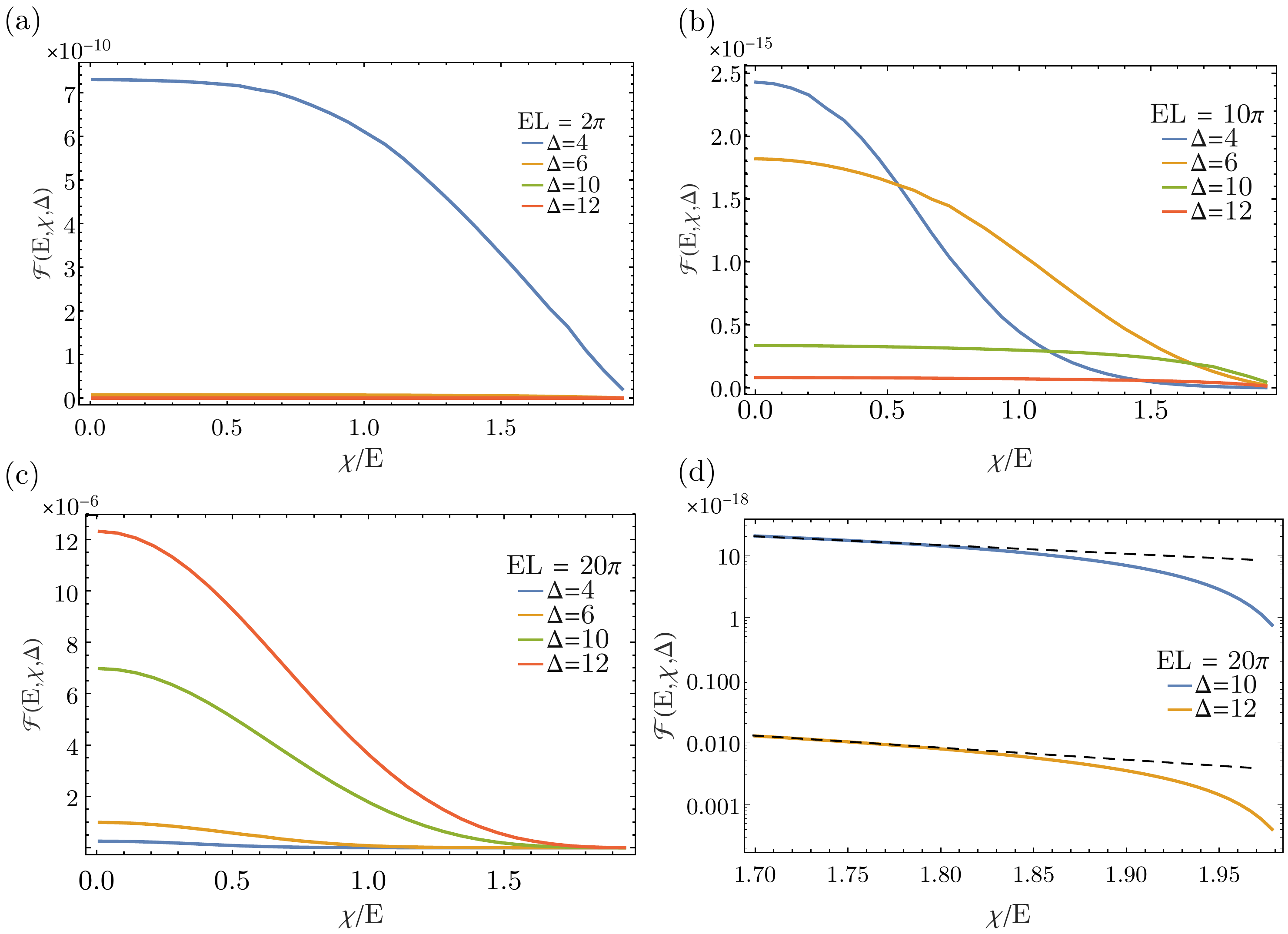}}
\vspace{-4mm}\caption{(a), (b) and (c) show the average of the square of matrix elements  $\mathcal{F}(E,\chi,\Delta)$ of scalars primary of various dimensions $\Delta$ as a function of the energy difference of the states. Computed numerically from eqs. \eqref{eq:F_definition}, \eqref{eq:Fhat_line} and \eqref{eq:DoS} with $c=20$. (d) shows the tails of $\mathcal{F}(E,\chi,\Delta)$ near $\chi\sim 2E$ in  logarithmic-linear scale. Dashed lines are a guide to they eye suggesting  the exponential suppression observed in the context of ETH in  one-dimensional systems \cite{Beugeling2015,Mondaini2017}.}
	\label{fig:off-diag}
\end{figure}

We now turn our attention to the matrix elements for states of different energies. Note that all our expressions are even in $\chi$, as % they should be since
everything is symmetric under interchange of $a$ and $b$. Furthermore,
$\chi \leq 2E$ by definition. Evaluating the explicit expression
  for $\mathcal{F}(E,\chi,\Delta)$ from eqs. \eqref{eq:F_definition}, \eqref{eq:Fhat_line} and \eqref{eq:DoS}
  numerically, there are three regimes of interest: $E\ll
\frac{\Delta^2}{c}$, $E \sim {\Delta^2}{c}$ and $E \gg
\frac{\Delta^2}{c}$. These three regimes are shown in figures \ref{fig:off-diag} (a), (b) and (c), respectively. The general characteristic for
  all three regimes is a maximum around $\chi=0$, followed by an
  exponential suppression near $\chi\sim 2E$ indicated by the dashed lines in Fig. \ref{fig:off-diag}-(d). This dependence on $\chi=E_a-E_b$ in
  the variance function $f(E_a,E_b)$ is what characteristically
  distinguishes ETH from random matrix theory (see
  e.g. \cite{Sonner:2017hxc}), and such exponential behavior in
  particular has been seen in numerical simulations
  of ETH systems \cite{Beugeling2015,Mondaini2017}.

\section{Probing a progressively disentangled geometry behind the horizon}
Armed with an expression for $\mathcal{F}(E,\chi,\Delta)$ (contained
in equations \eqref{eq:F_definition}, \eqref{eq:Fhat_line} and
\eqref{eq:DoS}), we can now compute Wightman functions of $\OO$ in
double-field states that are {\em arbitrarily} entangled, i.e. states
of the form
\begin{equation}
|\Psi\rangle = \frac{1}{\sqrt{Z_{\Psi}}} \sum_a \Psi(E_a) |a\rangle |a\rangle~,
\end{equation}
where $\Psi(E_a)$ is an arbitrary function of $E_a$, subject only to the constraint that the state have finite norm:
\begin{equation}
Z_\Psi=\sum_{a} |\Psi(E_a)|^2 < \infty~.
\end{equation}
The Wightman function is then simply
\begin{equation}
\langle \Psi|\OO(t) \otimes \OO(0)|\Psi \rangle=\frac{1}{Z_\psi} \int\limits_{-\infty}^\infty \dd\chi e^{-i\chi t} \int\limits_{|\chi|/2}^\infty \dd E ~\Psi^{*}\left(E{+}\frac{\chi}{2}\right) \Psi\left(E{-}\frac{\chi}{2}\right) \rho\left(E{+}\frac{\chi}{2}\right) \rho\left(E{-}\frac{\chi}{2}\right) \mathcal{F}(E,\chi,\Delta)~.
\end{equation}
The function $\Psi(E)$ gives us control over the entanglement between
the two copies of the CFT. In particular $\Psi(E)=e^{-\beta
  E/2}$ corresponds to the usual Boltzman distribution that leads to
the TFD state, and $\Psi(E)=\delta_{E,0}$ corresponds to two decoupled CFTs in
the ground state.

\begin{figure}[H]%
	  \begin{subfigure}{0.5\textwidth}
    \centering{  \hspace{-5mm}\includegraphics[scale=1.5]{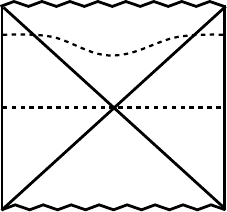}}
    %\caption{Caption text 1}
  \end{subfigure}
  \begin{subfigure}{0.5\textwidth}
    \centering{\includegraphics[scale=1.5]{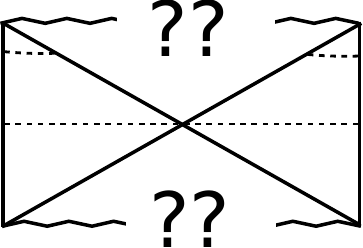}}
    %\caption{Caption text 2}
  \end{subfigure}
	\caption{Sketch of a Penrose diagram with dashed lines indicating the symmetric space-like geodesics for $t=0$ and $t\neq 0$. The diagram on the left is for the BTZ black hole, the one on the right for the state we are investigating.}
	\label{penrose}
\end{figure}

Our ability to compute Wightman functions in
  arbitrary entangled doubled-CFTs gives us a unique probe to study
  the spacetime behind the horizon, and its emergence/dissolution due to
  (dis)entanglement. In the thermally entangled TFD state the
  (logarithm of the)
  thermo-field double Wightman function for large $\Delta$ computes the length of geodesics between the
  two sides of the Kruskal extended BTZ black hole. For small
  deviations from the thermal state, a geodesic interpretation should
  still hold, but clearly for large deviations any geometrical notion
  should break down. This is clear both from the gedankenexperiment by
  Van Raamsdonk \cite{VanRaamsdonk:2010pw} quoted in the introduction and the simple observation
  that at $T=0$ the two copies of the CFT should be disentangled and
  hence disconnected --- no geodesic exists. According to the AdS/CFT
  correspondence, the quantum extension of
  this family of geodesics is precisely given by the Wightman function
  above for arbitrary entanglement, i.e. for arbitrary double-sided
  quantum geometries. Choosing different entanglement functions
  $\Psi(E_a)$ thus allows us to study the disentanglement of the
  spacetime behind the horizon, see Fig.\ref{penrose}. Note
    that the  Feynman two-point function for large $\Delta$ is dual to
    geodesics ending on the same boundary. These should be smooth, no
    matter the function $\Psi(E_a)$, as is clear from taking $\Psi(E_a)=\delta_{E_a,0}$.

We will choose here a rather drastic function $\Psi(E)=e^{-\beta E/2}
\Theta(\Lambda-E)$ that interpolates
discontinuously between the entangled thermofield double state and
un-entangled vacuum beyond the cutoff $\Lambda$. Thus we consider the
family of doubled states
\begin{equation}
|\Lambda\rangle \equiv \frac{1}{\sqrt{Z_\Lambda}} \sum_{a | E_a \leq \Lambda} e^{-\beta E_a/2} |a\rangle |a\rangle~,
\qquad Z_\Lambda \equiv  \sum_{a | E_a \leq \Lambda} e^{-\beta E_a}=\int_0^\Lambda dE \rho(E) e^{-\beta E}~.
\end{equation}

\begin{figure}[H]%
	\centering
	\resizebox{0.6\textwidth}{!}{\includegraphics{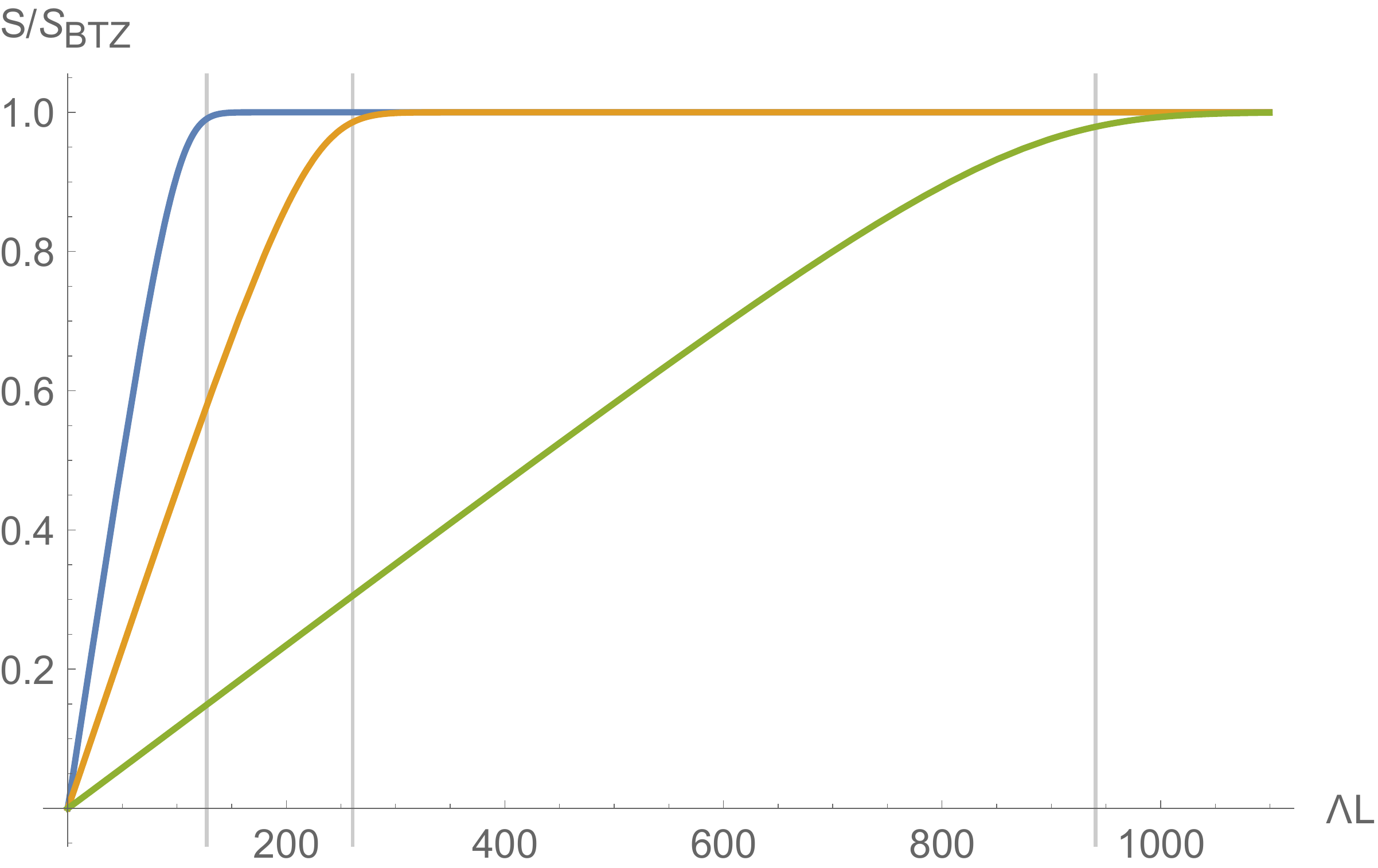}}
	\caption{Ratio of entanglement entropy of state with cutoff $\Lambda$ to BTZ entanglement entropy of the same $\beta$. $c=100$. The blue curve is $\beta/L=3/4$, the orange $\beta/L=1/2$ and the green $\beta/L=1/4$. The vertical lines correspond to a value of the cutoff where a saddle-point approximation predicts that the ratio should saturate.}
	\label{SSBTZ}
\end{figure}

By construction these states have less entanglement between the two sides than the thermofield double at inverse temperature $\beta$. To make this precise, we can compute the entanglement entropy between the two sides. It is given by
\begin{equation}
S=\beta \langle E \rangle_\Lambda ~~,~~ \langle E \rangle_{\Lambda}=-\frac{d}{d\beta} \log Z_\Lambda~=\frac{1}{Z_\Lambda} \int_0^\Lambda dE \rho(E) E e^{-\beta E}~.
\end{equation}
Comparing to the entropy of the thermal state (the BTZ black
hole) this gives
\begin{equation}
\frac{S}{S_{\text{BTZ}}}= \frac{\partial_\beta \log Z_\Lambda}{\partial_\beta \log Z(\beta)} = -\frac{6\beta^2}{\pi c L} \partial_\beta \log Z_\Lambda~.
\end{equation}
This ratio $S/S_{\text{BTZ}}$ starts at zero and grows linearly as a function of the cutoff $\Lambda$ until it saturates to 1, as seen in figure \ref{SSBTZ}. Thus, varying $\Lambda$ lets us directly dial the entanglement between the two copies of the CFT. The linearity and the saturation point can easily be understood using a saddle-point approximation. In the Cardy regime, $\rho(E) e^{-\beta E} \sim e^{-\beta E+ 2\sqrt{\frac{\pi c}{6}EL}}$ is strongly peaked around $E=\frac{\pi c}{6}\frac{L}{\beta^2}$. The width of this peak is $\frac{1}{\beta} \sqrt{\frac{\pi c L}{3}}$. Therefore, we expect the average energy to match closely the thermal answer when the range of integration includes most of the peak, i.e when $\Lambda \geq \frac{\pi c}{6} \frac{L}{\beta^2} + \frac{1}{\beta}\sqrt{\frac{\pi c L}{3}} $. This is indeed what we observe. The linearity can be explained by the fact that for $E\ll\frac{\pi c L}{6\beta^2}$ the integral is dominated by the value of the integrand at $E=\Lambda$, and so $\langle E \rangle_\Lambda \sim \Lambda$. Note that in the large $c$ regime the relative width of the peak goes to zero since it scales as $c^{-1/2}$.

We now use the Wightman function in the modified state $|\Lambda\rangle$, based on the entanglement function $\Psi(E)=e^{-\beta E/2}
\Theta(\Lambda-E)$,
\begin{align}
G^\Lambda(t)\equiv &
\frac{1}{Z_\Lambda} \int\limits_{-\Lambda}^{\Lambda} \dd \chi \int\limits_{\chi/2}^{\Lambda-{\chi/ 2}} \dd E\  e^{-i\chi t}e^{-\beta E}\rho\left(E+{\chi\over 2}\right)\rho\left(E-{\chi\over 2}\right)\mathcal{F}(E,\chi,\Delta)\,\label{eq:G_cutoff}~,
\end{align}
to study what happens to the geometry as we vary the entanglement
between the two boundaries.

The leading behavior for large $\Delta$ is related to the length
  of  geodesics connecting the points $(\phi,t/2)$ on both boundaries
  \footnote{There is a symmetry that lets us shift the time of the
    insertions on both boundaries by the same amount in opposite
    direction. This corresponds to the action of the usual timelike
    Killing vector in the black hole geometry. Note that this symmetry is preserved in our modified state.} in
  the bulk dual of our modified state. 
The distance between the two boundaries is of course formally
infinite. To regulate we compare to the geodesics computed in the
thermal state dual to the BTZ black hole, i.e. we compute 
\begin{equation}
\mathcal{L}_{\Lambda}-\mathcal{L_{\text{BTZ}}} \sim -\frac{1}{\Delta} \log\frac{G^\Lambda (t)}{G^\beta (t)}  ~.
\end{equation}

\begin{figure}[H]
  \hspace{-2mm}\begin{subfigure}{0.5\textwidth}
    \centering{  \includegraphics[scale=0.54]{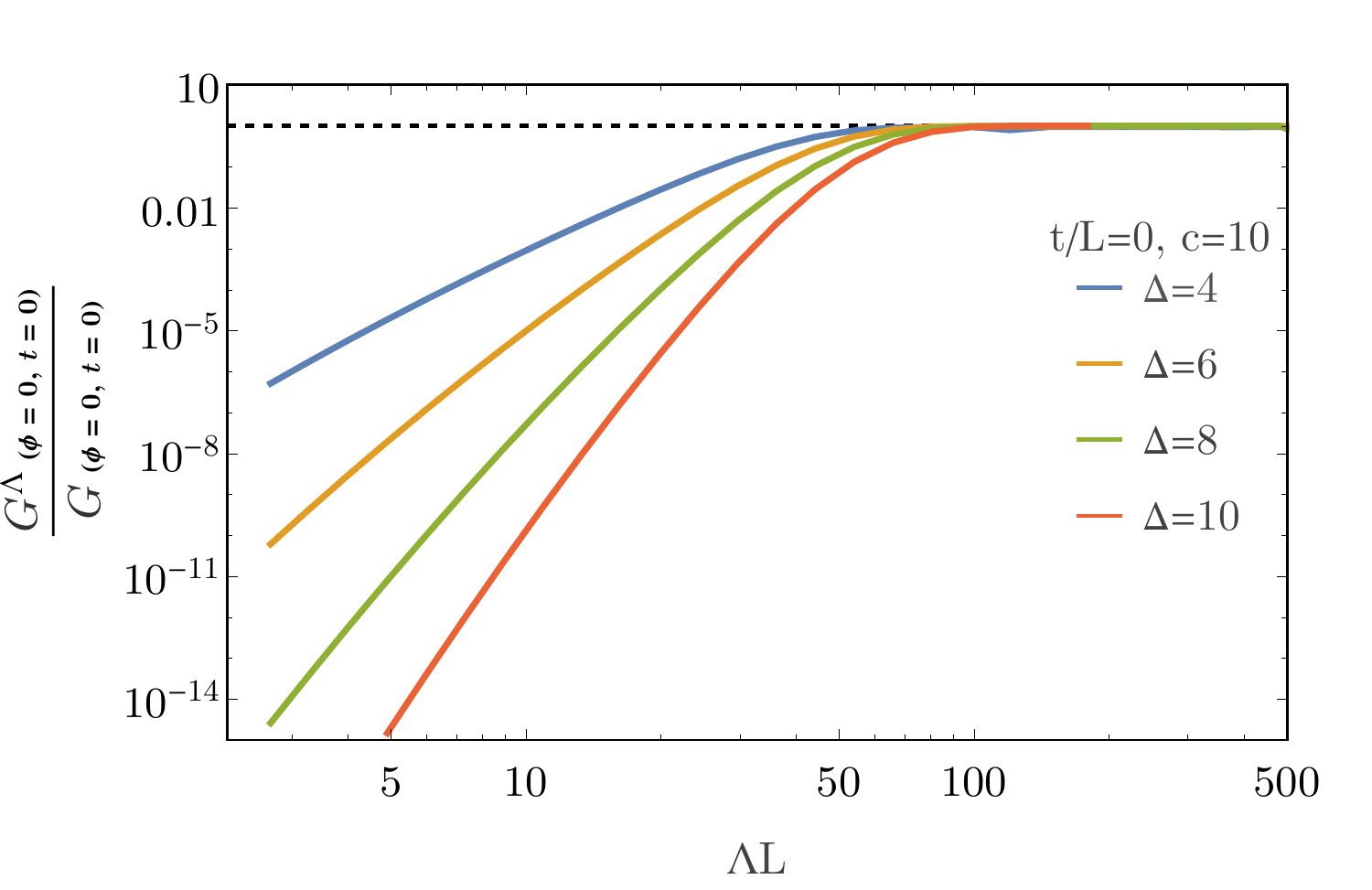}}
    %\caption{Caption text 1}
  \end{subfigure}\hspace{2mm}
  \begin{subfigure}{0.5\textwidth}
    \centering{\includegraphics[scale=0.48]{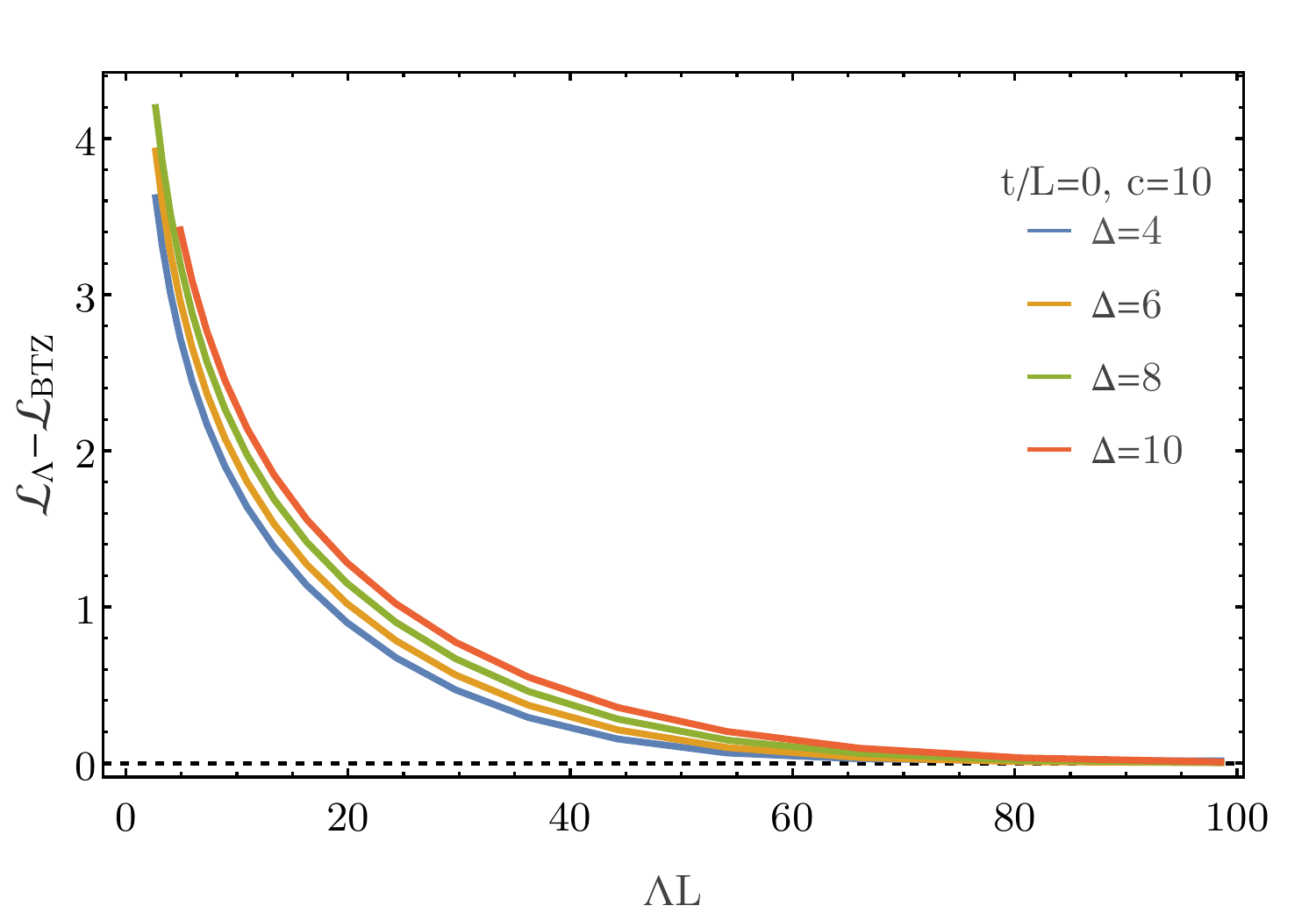}}
    %\caption{Caption text 2}
  \end{subfigure}
  \caption{Normalized Wightman function with finite cutoff (left) and length difference from BTZ (right) as defined in eq. \eqref{eq:G_cutoff} for fixed $\beta/L=1/2$. The central charge is $c=10$ and space separation is zero.}
	\label{fig:G_cutoff}
\end{figure}

We start our analysis at $t=0$. Figure \ref{fig:G_cutoff} shows that
as $\Lambda \rightarrow \infty$, we smoothly recover the black
hole. The $t=0$ geodesics are special in that they do not cross
the horizon: they connect the two boundaries through the bifurcation
surface. This indicates that as we decrease the entanglement between
the two boundaries, the two patches move apart (as expected) but are
not otherwise affected. In particular, we can use the Killing symmetry
to move the endpoints of the geodesic up on one side and down on the
other without affecting the answer. This corresponds to sweeping the
entire geometry outside of the horizon. To see this,
  consider a point outside the horizon. This point  always lies on a
  spacelike geodesic connecting a boundary point at some time (say
  $t_0$) and the bifurcation
  surface. 
By symmetry, that geodesic will always intersect the other boundary at
Schwarzschild time $-t_0-i\frac{\beta}{2}$, i.e boundary time $t_0$. 
As also remarked before, one sees that the bulk geometry dual to our modified 
state is not dramatically affected outside the horizon by the decrease in entanglement. 
In particular, we know that as $\Lambda\rightarrow 0$, we simply 
recover two disconnected patches of AdS.

\begin{figure}[H]
 \hspace{-3mm} \begin{subfigure}{0.5\textwidth}
    \centering{  \includegraphics[scale=0.53]{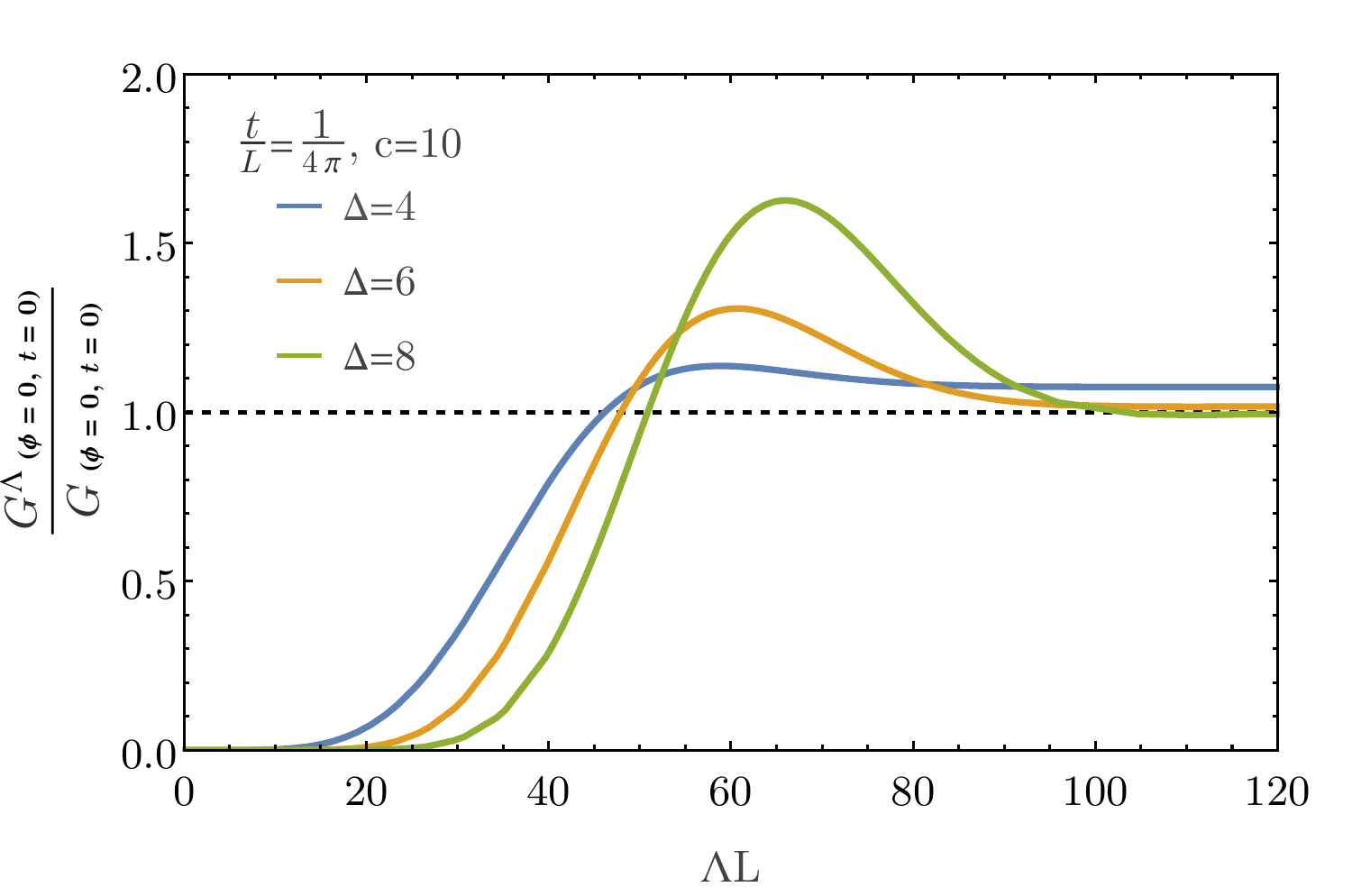}}
    %\caption{Caption text 1}
  \end{subfigure}\hspace{2mm}
  \begin{subfigure}{0.5\textwidth}
    \centering{\includegraphics[scale=0.5]{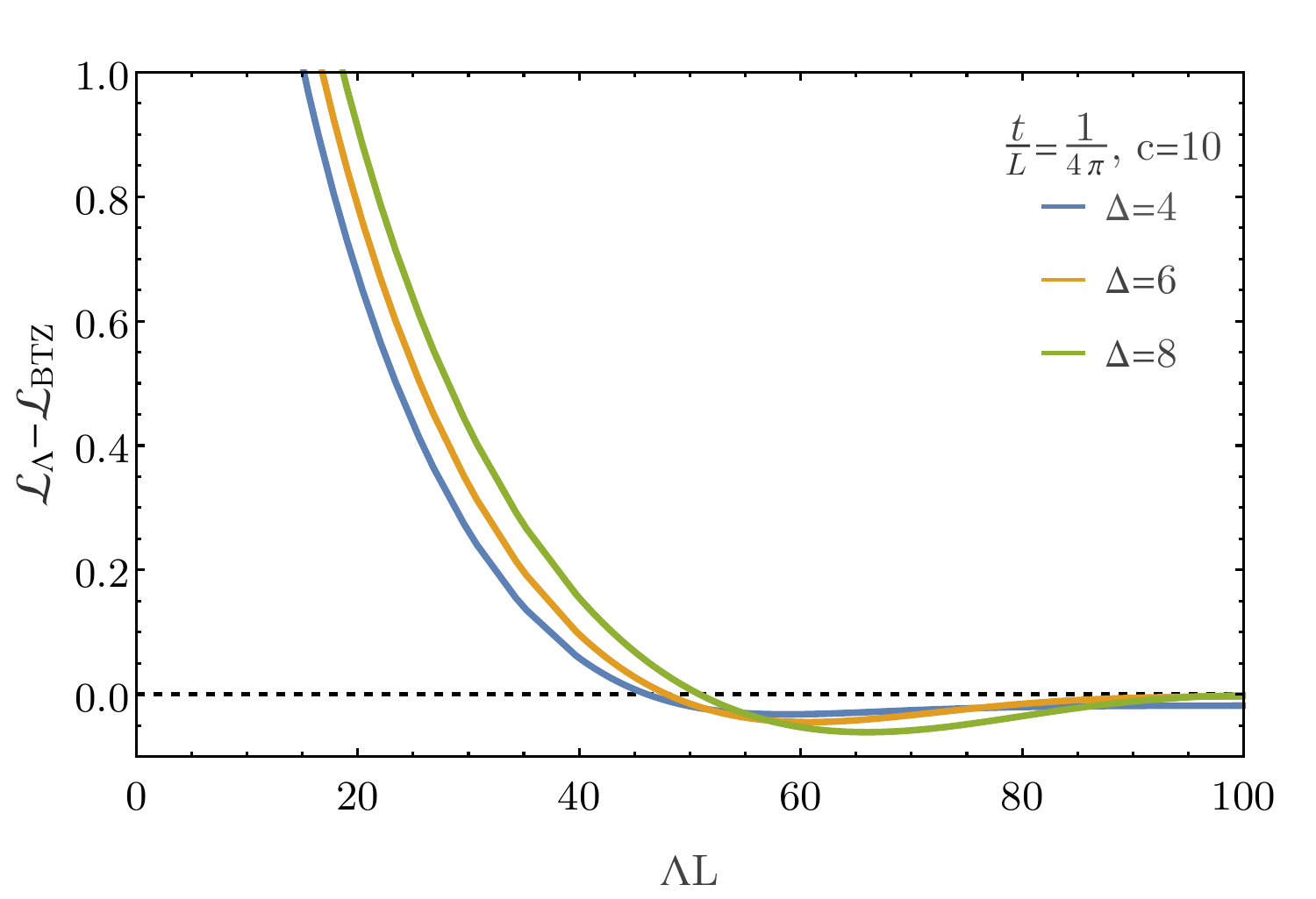}}
    %\caption{Caption text 2}
  \end{subfigure}
\caption{Normalized Wightman function with finite cutoff (left) and length difference from BTZ (right) as defined in eq. \eqref{eq:G_cutoff} for fixed $\beta/L=1/2$. The central charge is $c=10$ and spatial separation is zero.}\label{fig:GLambda_t_vs_beta}
\end{figure}

\begin{figure}[H]
  \hspace{-3mm}\begin{subfigure}{0.5\textwidth}
    \centering{  \includegraphics[scale=0.53]{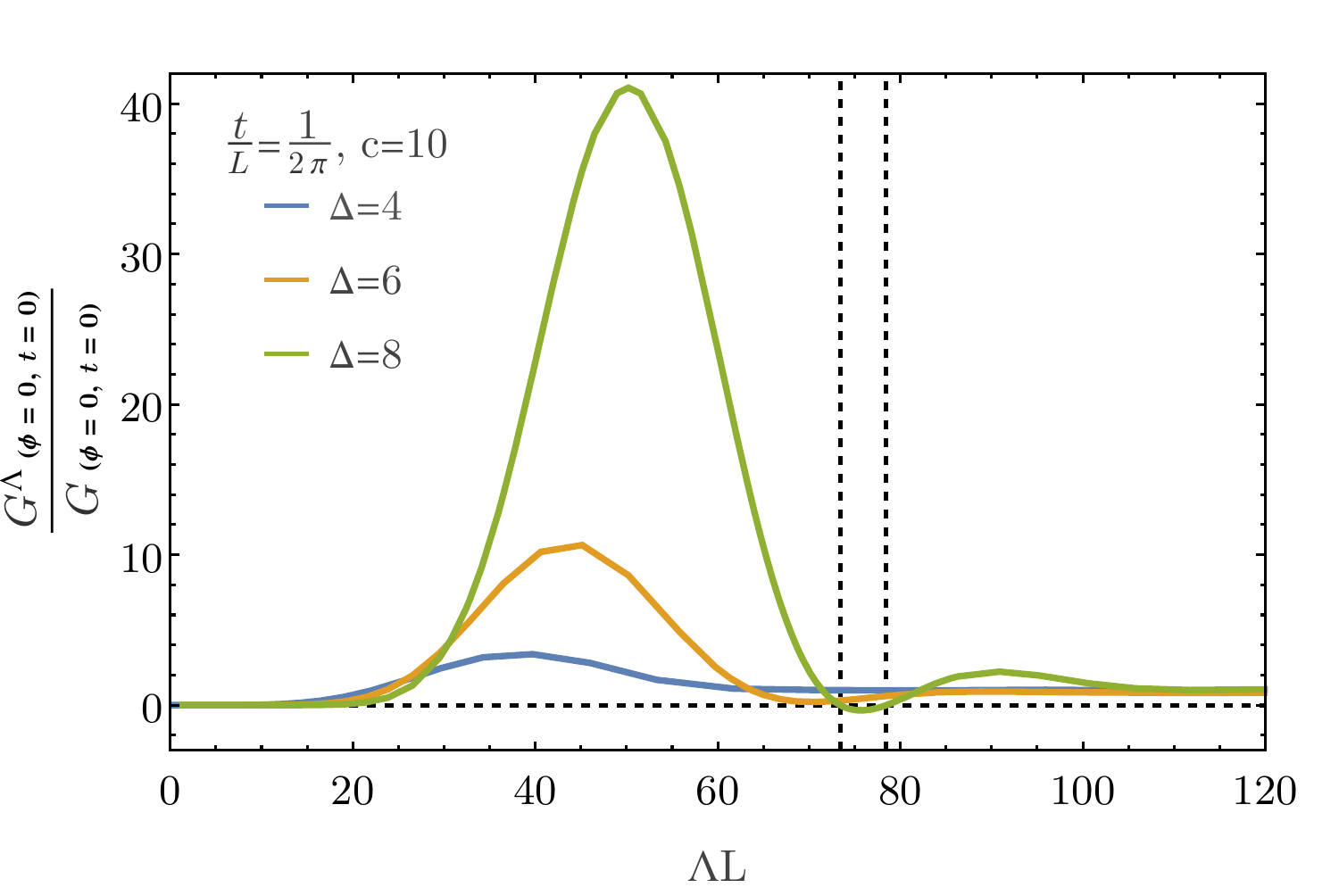}}
    %\caption{Caption text 1}
  \end{subfigure}\hspace{1mm}
  \begin{subfigure}{0.5\textwidth}
    \centering{\includegraphics[scale=0.53]{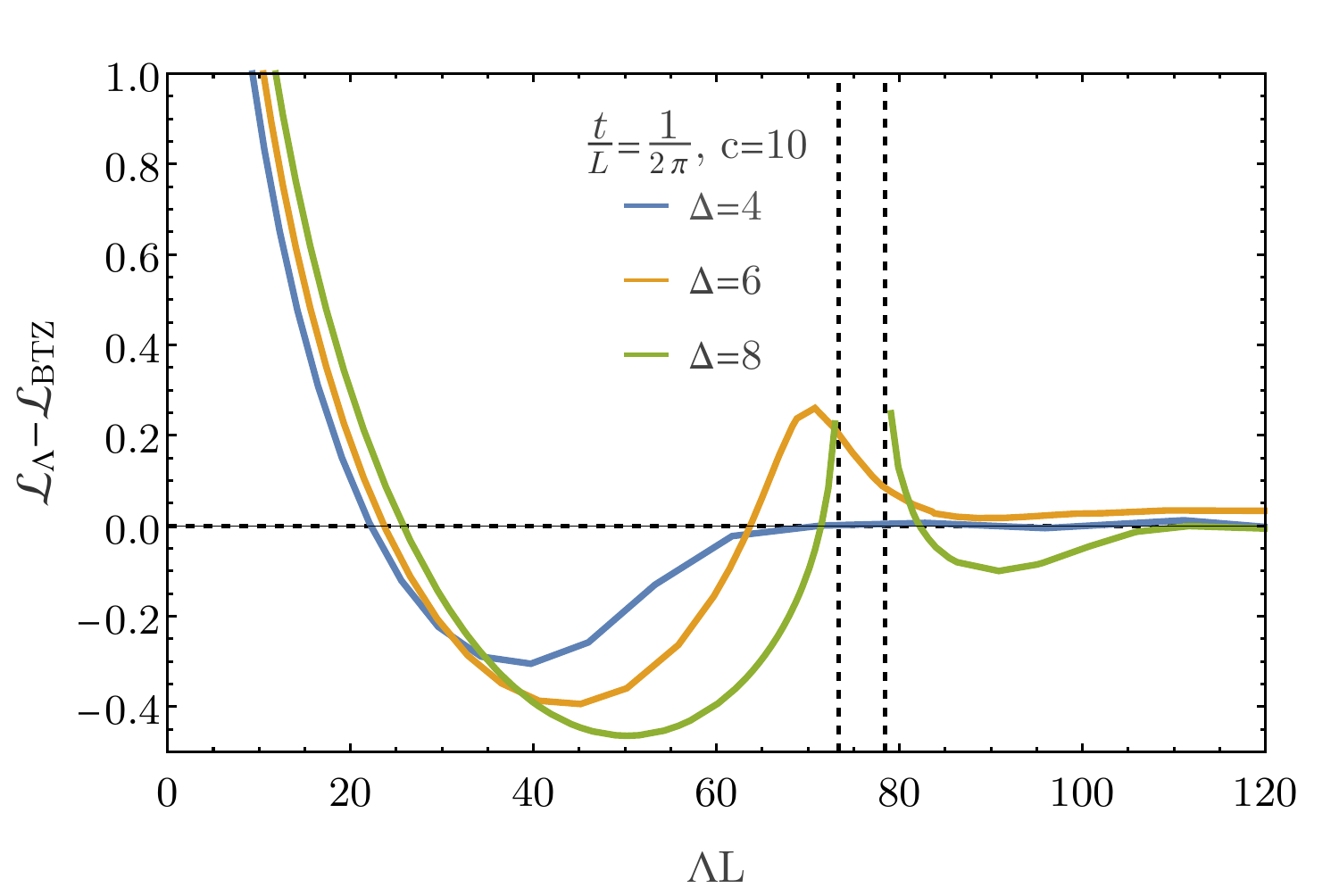}}
    %\caption{Caption text 2}
  \end{subfigure}
	\caption{Normalized Wightman function with finite cutoff (left) and length difference from BTZ (right) as defined in eq. \eqref{eq:G_cutoff} for fixed $\beta/L=1/2$. The central charge is $c=10$ and spatial separation is zero. The vertical dashed lines indicate the  region where the Wightman function is negative for $\Delta=8$. Consequently, the geodesic length cannot be defined from the Wightman function.}\label{fig:GLambda_t2_vs_beta}

\end{figure}

As we look at finite $t\neq 0$, we start probing the geometry behind
the horizon. 
Now we experience the drastic effects of our severe change in
  entanglement. Figure \ref{fig:GLambda_t_vs_beta} shows what happens
for geodesics corresponding to early times. Instead of the ratio of
the two-point functions increasing monotonically as a function of
cutoff, it increases past one and then decreases again, eventually
saturating to one.  The intuition that the geometry behind the horizon
simply becomes more connected in a smooth fashion as $\Lambda
\rightarrow \infty$ -- which appears to be supported by the $t=0$
geodesics through the bifurcation surface -- is clearly wrong! This
becomes even clearer as we increase the time, which corresponds to
geodesics in the black hole background having a greater fraction of
their length behind the horizon. Figure \ref{fig:GLambda_t2_vs_beta}
shows that the changes in the ratio of the Wightman
functions as a function of $\Lambda L$ increase for larger
  $t$. The Wightman function $G^\Lambda(t)$ actually becomes negative
for some values of $\Lambda$.\footnote{There is nothing from the field
  theory perspective that would prevent this: at finite $t$, the only
  constraint is that the Wightman function be real.} This signals a
complete breakdown of the geodesic approximation. At best, one
  could argue that there are multiple geodesics that interfere
destructively, with the path integral whose saddlepoints give the
  geodesics very sensitive to small variations in $\Lambda$. 
This cannot be expected to describe a classical space-time. 
Note that since the time shift symmetry is manifestly preserved in the
expression for the Wightman function for any value of $\Lambda$, we cannot think of the geometric breakdown behind the horizon as being localized in one particular region.

\begin{figure}[H]
  \begin{subfigure}{0.5\textwidth}
    \centering{  \hspace{-7mm}\includegraphics[scale=0.53]{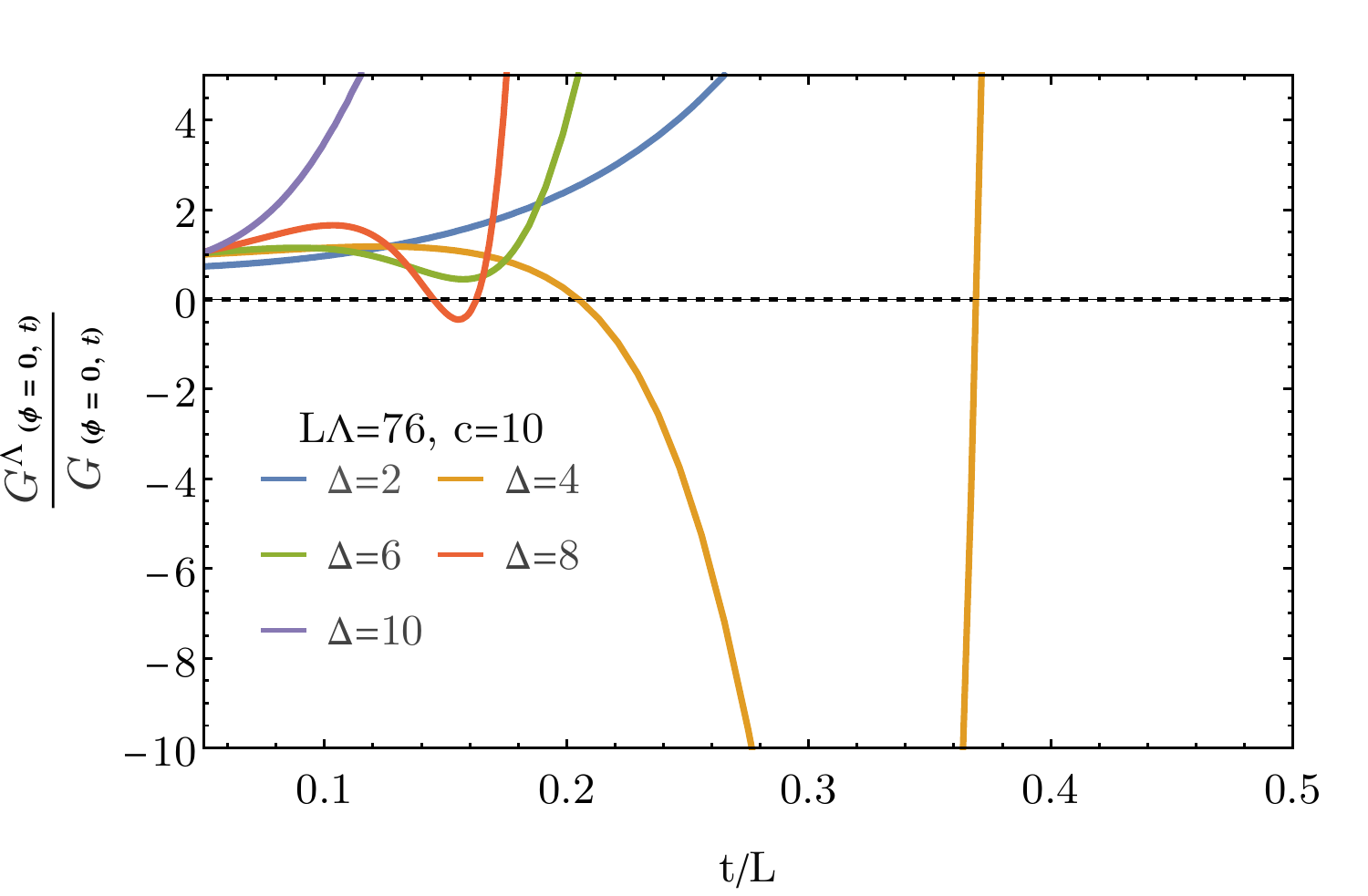}}
    %\caption{Caption text 1}
  \end{subfigure}\hspace{-3mm}
  \begin{subfigure}{0.5\textwidth}
    \centering{\includegraphics[scale=0.515]{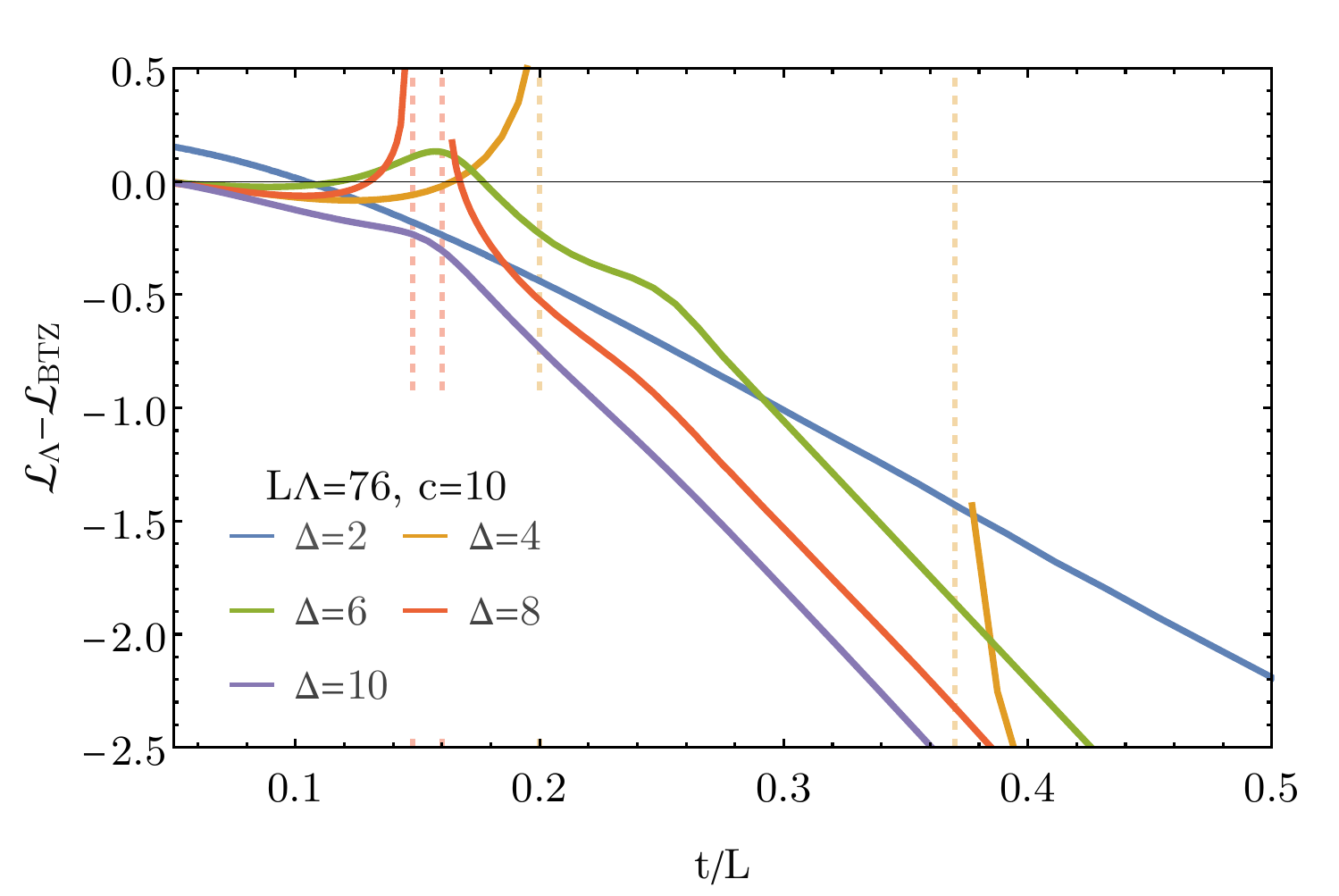}}
    %\caption{Caption text 2}
  \end{subfigure}
	\caption{Normalized Wightman function with finite cutoff (left) and length difference from BTZ (right) as a function of time for fixed $L\Lambda=76$, $\beta/L=1/2$, $L=2\pi$ and $c=10$. The space separation is zero.}\label{fig:GLambda2}

\end{figure}

Rather than the cut-off dependence for fixed $t$, one can also look at the time dependence for fixed
cutoff.\footnote{Note that there is of course no expectation that the
  difference in lengths saturate as $t\rightarrow\infty$.} Doing so
dispels the notion that the non-geometric behaviour described
previously only occurs for some values of $\Delta$. For example, note
that Fig. \ref{fig:GLambda_t2_vs_beta} shows that for $t/L=1/2\pi$,
$\Lambda L\sim 76$ and $\Delta=4$, the Wightman function is
positive. However, we see from Fig. \ref{fig:GLambda2} that for the
same cutoff and $ 0.20\lesssim t/L\lesssim 0.35$ the Wightman
function for $\Delta=4$ does become negative.

Clearly, when the two copies of the CFT are too disentangled,
  no conventional notion of spacetime survives. Vice versa, by dialing
  the entanglement we can now literally see classical spacetime
  emerge/dissolve in a rather precise way. 
A geometric notion may
  survive for very large but finite $\Lambda$. Indeed, for all values of $t$ (including $t=0$), there is a value of $\Lambda$ past which the Wightman function essentially matches the thermal answer. This value of the cutoff increases with $c$. This was to be expected, since the total energy of the black hole also increases in $c$ and we certainly do not expect a classical bulk for $\Lambda \ll \langle E \rangle_\text{BTZ}$. Given the strongly peaked nature of $\rho(E)e^{-\beta E}$ around $E= \frac{\pi c L}{6\beta^2}$ and of $\mathcal{F}(E,\chi,\Delta)$ around $\chi=0$, we expect large fluctuations around an energy of order $E \sim \frac{c L}{\beta^2}$, within a window of width scaling as $\frac{\sqrt{c L}}{\beta}$. Our numerical results are consistent with this.

\section{Summary and discussion}
In this article, we have achieved two things. First, we have derived
an expression for the average of the square of matrix elements of a
scalar primary of integer dimension taken between states of different
energies. This is the function we call
$\mathcal{F}(E,\chi,\Delta)$. We have obtained an exact, formal
expression for this function valid for any holographic CFT at
finite temperature and radius (in the deconfined phase). A more
useful, Cardy-like formula, was also found. If interpreted in the
context of the eigenstate thermalization hypothesis,
$\mathcal{F}(E,\chi,\Delta)$ computes the square of the universal part
of the variance, $f(E,\chi)$ for holographic CFTs. 

Secondly, we have demonstrated that the knowledge of this function can be used to make strong statements about the dual bulk geometry. More precisely, it allows us to compute the Wightman function for states with an arbitrary energy-dependent entanglement pattern between the two copies of the theory. We have focused on the most brutal possible modification to the thermofield double state, chopping off the contribution of all energy levels above some cutoff. This allowed us to dial the entanglement from a thermal state to two disentangled copies of the vacuum. We have seen explicitly that below some amount of entanglement (which scales with $c$), no clear notion of a semi-classical geometry behind the horizon can exist.

There are several possible extensions to this work. First, let us note that the space of entanglement patterns we can explore is now essentially the space of square-normalizable functions. An obvious generalization of our state $|\Lambda\rangle$ is to remove states within an energy window. Another interesting avenue to explore may be how this function can help understand traversable wormholes \cite{Gao:2016bin} or dynamical space-times.

\acknowledgments

We thank Alex Belin, Enrico Brehm, Diptarka Das, Shouvik Datta, Anatoly Dymasky, Tom Hartman, Nima Lashkari, Alex Maloney, Charles Rabideau, Julian Sonner, and  Jan Zaanen for discussions. This research was supported in part by a VICI award of the Netherlands Organization for Scientific Research (NWO), by the Netherlands Organization for Scientific Research/Ministry of Science and Education (NWO/OCW), and by the Foundation for Research into Fundamental Matter (FOM).

\appendix
\section{Spatial dependence}\label{app1}

Here we generalize our formal result from Eq. \eqref{eq:formal} for operators inserted at different spatial coordinate $\phi$. In this case, Eq. \eqref{eq:TFD_correlator_2} is
\begin{align}
G^{\beta}(t_{12},\phi_{12})&\equiv \tfdl \OO(t_1,\phi_1) \otimes \OO(t_2,\phi_2) \tfdr = \frac{1}{Z(\beta)} \text{Tr}(\rho_\beta \OO(t_1-i\beta/2,\phi_1) \OO(t_2,\phi_2))\nonumber\\
&=\frac{1}{Z(\beta)} \sum_{a,b} e^{-\beta {E_a+E_b\over 2}} e^{i (E_a-E_b)t_{12}} \langle a |\OO(\phi_1)  |b \rangle \langle a |\OO(\phi_2)  |b \rangle\\
&=\frac{1}{Z(\beta)} \sum_{E_a, E_b}  e^{-\beta {E_a+E_b\over 2}}e^{i (E_a-E_b)t_{12}} e^{S(E_a)} e^{S(E_b)} \mathcal{F}(E_a,E_b,\phi_{12},\Delta)~.
\end{align}
where $\phi_{12}\equiv \phi_1-\phi_2$ and $t_{12}\equiv t_1-t_2$. As in the case of operators inserted at the same $\phi$, we have split the sum into a sum over energies and defined $\mathcal{F}(E_a,E_b,\Delta)$ in the same way, except that it  contains an additional spatial depencende in $\phi_{12}$.  The calculation follows in exactly the same way as explained in Sec. \ref{sec:Setup}. Namely, we compute $\mathcal{F}(E_a,E_b,\phi_{12},\Delta)$ by inverting
\begin{align}
G^{\beta}(t_{12},\phi_{12})&= \frac{e^{-\beta E_L}}{Z(\beta)}
\int\limits_{-\infty}^{\infty} \dd \chi e^{-i\chi t} 
\int\limits_{0}^{\infty} \dd Ee^{-\beta E}
\rho\left(E+{\chi\over 2}\right)\rho\left(E-{\chi\over 2}\right) 
 \Theta\left(E{-}{|\chi|\over 2}\right)\mathcal{F}(E,\chi,\phi_{12},\Delta)~,
\end{align}
where
\begin{equation}
G^{\beta}(t_{12},\phi_{12})=\sum_{n\in \mathbb{Z}} C_\OO \frac{(2\pi/\beta)^{2\Delta}}{\left[\cosh \left(2\pi/\beta(\phi_{12}+n L)\right)+\cosh\left(2\pi (t_{12})/\beta\right)\right]^\Delta}\,.
\end{equation}
The inversion of the Fourier transform has already been given in Sec. \ref{sec:FT}, eqs. \eqref{eq:formal}-\eqref{eq:inverseFT}, where now $b_n(\phi_{12})=\cosh(2\pi/\beta(\phi_{12}+nL))$ instead. Similarly, the inversion of the Laplace transform carries through as explained in Sec. \ref{sec:LT}. The final result is therefore,
\begin{equation}
\rho\left(E+{\chi\over 2}\right)\rho\left(E-{\chi\over 2}\right) 
 \Theta\left(E{-}{|\chi|\over 2}\right)\mathcal{F}(E,\chi,\phi_{12},\Delta) = 
 \frac{C_\OO}{{2\pi}}\int\limits_{0}^{E} \dd E' \rho(E'{+}E_L) \sum_{n \in \mathbb{Z}}\hat{F}_n(E-E',\chi,\Delta)~.
\end{equation}
where $\hat{F}_n(E-E',\chi,\Delta)$, given in Eq. \eqref{eq:Fnhat}, now depends on $\phi_{12}$ through $b_n(\phi_{12})=\cosh(2\pi/\beta(\phi_{12}+nL))$.
Note that $\mathcal{F}(E,\chi,\phi_{12},\Delta)$ contains information about more than the spectrum of the CFT, as is necessary to reconstruct Wightman functions for operators inserted at different points in space.

\section{Comments on corrections to the density of sates}\label{app:DoS}
In Sec. \ref{sec:LT} we have obtained the density of states using the leading-in $c$ partition function. We restricted to the region $E-{\pi c\over 6L}>0$ so that we could ignore the subleading corrections in $c$ to the partition function. Here, we explain how these non-universal corrections determine the density for $E-{\pi c\over 6L}<0$.

The partition function to leading order in $c$ is \cite{Hartman:2014oaa}
\begin{align}
\log Z=
\begin{cases}
& {\pi c L\over 6\beta} +O(c^0) ~~~,~~\beta \leq 2\pi~,\\
& {\beta c\pi \over 6L} +O(c^0) ~~~,~~\beta \geq 2\pi,\\
\end{cases}
\end{align}
which can be parametrised as $Z(\beta) = f(\beta)e^{\pi c L\over 6\beta}+f(L^2/\beta)e^{\beta c\pi \over 6L} $, where $f$ contains all the subleading corrections. This parametrisation is chosen so that modular invariance is evident. We are now in a position to see  that the density of states for $E<{\pi c\over 6L}$ is obtained from the non-universal corrections to the partition function:
\begin{align}
Z(\beta) &= \int\limits_{-{\pi c\over 6L}}^\infty \dd E \rho(E) e^{\beta E}= e^{{\beta\pi c\over 6L}}\int\limits_{0}^\infty \dd E \rho(E) e^{\beta E}\implies \notag\\
\rho(E) &= {1\over 2\pi i}\int\limits_{\gamma_0+i \mathbb{R}} \dd \beta e^{\beta\left(E-{\pi c\over 6L}\right)}Z(\beta)\notag\\
&= 
{1\over 2\pi i}\int\limits_{\gamma_0+i \mathbb{R}} \dd \beta f(\beta)e^{\beta\left(E-{\pi c\over 6L}\right)+{\pi c L\over 6\beta}}+
{1\over 2\pi i}\int\limits_{\gamma_0+i \mathbb{R}} \dd \beta f(L^2/\beta)e^{\beta\left(E-{\pi c\over 6L}\right)+{\beta c\pi \over 6L} }\notag \\
&= 
{1\over 2\pi i}\int\limits_{\gamma_0+i \mathbb{R}} \dd \beta f(\beta)e^{\beta\left(E-{\pi c\over 6L}\right)+{\pi c L\over 6\beta}}+
{1\over 2\pi i}\int\limits_{\gamma_0+i \mathbb{R}} \dd \beta f(L^2/\beta)e^{\beta E}\,,\notag
\end{align}
where, as before, $\gamma_0$ is fixed so that the integration line is set to the right of all singularities of the integrand. 

For $E>{\pi c\over 6L}$ the first integral is obtained by closing the contour to the left (Bromwich contour), while the second integral is obtained by closing the contour to the right giving zero contribution to the density of states. On the other hand for  $E<{\pi c\over 6L}$, the situation is reversed and the density of states is obtained from the second integral with the contour closed to the left. Notice that, due to modular  invariance, the corrections $f(\beta)$ are needed non-perturbatively in $\beta$. 

\bibliographystyle{JHEP}
\bibliography{library}

\end{document}